\documentclass[aps,pra,twocolumn,groupedaddress,showpacs]{revtex4-1}
\usepackage{graphicx}
\usepackage{amsmath}
\usepackage{textcomp}
\usepackage{mathrsfs}
\usepackage{amsfonts}
\usepackage{mathtools}
\usepackage{tikz} 
\usetikzlibrary{positioning}
\usetikzlibrary{shapes,arrows,positioning,automata,backgrounds,calc,er,patterns}
\usepackage{tikz-feynman}
\tikzfeynmanset{compat=1.0.0}
\usepackage{dirtytalk}

\DeclareMathOperator{\sinc}{sinc}
\DeclareMathOperator{\sech}{sech}

\begin{document}

\newcommand{\vett}[1]{\textbf{#1}}
\newcommand{\uvett}[1]{\hat{\textbf{#1}}}
\newcommand{\fieldE}[2]{E(\vett{#1},#2)}
\newcommand{\fieldA}[2]{A(\vett{#1},#2)}

\newcommand{\fieldX}[3]{\psi_{#1,#2}^{(#3)}(R,\zeta)}
\newcommand{\fieldXc}[3]{\psi_{#1,#2}^{(#3)*}(R,\zeta)}

\newcommand{\fieldXz}[3]{\psi_{#1,#2}^{(#3)}(R,z)}
\newcommand{\fieldXcz}[3]{\psi_{#1,#2}^{(#3)*}(R,z)}

\newcommand{\fieldXo}[2]{\psi_{#1,#2}^{(0)}(r,X)}
\newcommand{\fieldXco}[2]{\psi_{#1,#2}^{(0)*}(r,X)}

\newcommand{\fieldXp}[3]{\psi_{#1,#2}^{(#3)}(R',\zeta)}

\newcommand{\fieldXa}[3]{\psi_{#1,#2}^{(#3)}(R,\zeta_1)}
\newcommand{\fieldXb}[3]{\psi_{#1,#2}^{(#3)}(R,\zeta_2)}

\newcommand{\creatX}[3]{\hat{a}_{#1,#2}^{\dagger}(#3)}
\newcommand{\annX}[3]{\hat{a}_{#1,#2}(#3)}
\newcommand{\creatXb}[3]{\hat{b}_{#1,#2}^{\dagger}(#3)}
\newcommand{\annXb}[3]{\hat{b}_{#1,#2}(#3)}
\newcommand{\annA}[2]{\hat{A}(\vett{#1},#2)}
\newcommand{\creatA}[2]{\hat{A}^{\dagger}(\vett{#1},#2)}

\newcommand{\beq}{\begin{equation}}
\newcommand{\eeq}{\end{equation}}
\newcommand{\barr}{\begin{eqnarray}}
\newcommand{\earr}{\end{eqnarray}}
\newcommand{\bseq}{\begin{subequations}}
\newcommand{\eseq}{\end{subequations}}
\newcommand{\bal}{\begin{align}}
\newcommand{\eal}{\end{align}}
\newcommand{\ket}[1]{|#1\rangle}
\newcommand{\bra}[1]{\langle #1|}
\newcommand{\expectation}[3]{\langle #1|#2|#3 \rangle}
\newcommand{\braket}[2]{\langle #1|#2\rangle}

\newcommand{\multi}{\{\vett{m}\}}
\newcommand{\multid}{\{\vett{a}\}}
\newcommand{\multit}{\{\vett{b}\}}

\newcommand{\annF}[3]{\hat{\phi}_{#1#2}(#3)}
\newcommand{\creatF}[3]{\hat{\phi}_{#1#2}^{\dagger}(#3)}
\newcommand{\annFd}[2]{\hat{\phi}_{#1#2}}
\newcommand{\creatFd}[2]{\hat{\phi}_{#1#2}^{\dagger}}

\newcommand{\creatXt}[3]{\hat{a}_{#1,#2}^{\dagger}(#3,t)}
\newcommand{\annXt}[3]{\hat{a}_{#1,#2}(#3,t)}
\newcommand{\annFt}[3]{\hat{\phi}_{#1#2}(#3,t)}
\newcommand{\creatFt}[3]{\hat{\phi}_{#1#2}^{\dagger}(#3,t)}

\title{A path integral description of quantum nonlinear optics in arbitrary media}

\author{Mosbah Difallah$^{1}$}
\author{Alexander Szameit$^{2}$}
\author{Marco Ornigotti$^{2,3}$}
\email{marco.ornigotti@tuni.fi}
\affiliation{$^1$ Department of Physics, Faculty of Exact Sciences, University of El Oued, 39000, El Oued, Algeria}
\affiliation{$^2$Institut f\"ur Physik, Universit\"at Rostock, Albert-Einstein-Stra\ss e 23, 18059 Rostock, Germany}
\affiliation{$^3$ Laboratory of Photonics, Physics Unit, Tampere University, Tampere, Finland}

\begin{abstract}
We present a method, based on Feynman path integrals, to describe the propagation and properties of the quantised electromagnetic field in an arbitrary, nonlinear medium. We provide a general theory, valid for any order of optical nonlinearity, and we then specialise the the case of second order nonlinear processes. In particular, we show, that second-order nonlinear processes in arbitrary media, under the undepleted pump approximation, can be described by an effective free electromagnetic field, propagating in a vacuum, dressed by the medium itself. Moreover, we show,that the probability of such processes to occur is related to the biphoton propagator, which contains informations about the structure of the medium, its nonlinear properties, and the structure of the pump beam.
\end{abstract}

\maketitle
\section{Introduction}
Since the early days of laser physics, nonlinear optics, i.e., the study of the nonlinear interaction of the electromagnetic field with matter, has been a very successful, and intriguing field of research. The possibility of creating electromagnetic waves oscillating at new frequencies, with respect to the one possessed by the impinging beam, for example, is at the basis of modern laser-based devices, such as supercontinuum light sources \cite{supercontinuum}. The study of nonlinear dynamics of light, moreover, contributed to give a lot of insight in the physics of photonic crystals \cite{solitonComm}, optical fibers \cite{rogue}, and discrete optical systems \cite{rogue2}.

Nowadays, with the advent of quantum technologies, nonlinear optics has become the vital part of any quantum optics experiment, as spontaneous parametric down-conversion (SPDC), i.e., the generation of a photon pair from a bright laser beam \cite{boyd}, constitutes the primary source of entangled photons \cite{loudon}. In the last years, a significant effort has been made, to incorporate these sources of entangled photons in integrated on-chip platform, with the ultimate goal of realising fully integrated quantum devices, which will, ultimately, constitute the basis for quantum computers \cite{quantumComputer}. To this aim, integration of nonlinear effects in different nanostructures, such as quantum dots \cite{quantumDot}, quantum wells \cite{quantumWell}, photonic waveguides \cite{waveguide1,waveguide2,waveguide3}, plasmonic structures \cite{plasmon2}, and metamaterials, have been thoroughly investigated. Metamaterials, in particular, have the advantage of offering a simple way to tailor and enhance nonlinear processes \cite{meta1,meta2,meta3}. Metamaterials, however, are quite complex structures, and understanding and controlling their nonlinear properties, is not a simple task. A full understanding of their nonlinear properties, as well as the possibility to understand and control their onset, however, is a key feature, towards the use of plasmonic and metamaterial nanostructures for fully integrated quantum devices. For these reasons, having at disposal numerical and analytical techniques, to correctly model the nonlinear interactions of the electromagnetic field in such structures, is of paramount importance, as it provides the fundamental guidelines for the design of efficient structures and devices. The majority of theoretical frameworks currently available, however, has been developed for lossless systems (such as, for example, optical waveguides), systems made of dispersionless elements only \cite{dielectric2,dielectric3,dielectric4}, or complex geometries, but only interacting with few optical modes \cite{dielectric5,dielectric6,dielectric7}. An exception to this is represented by a recent work, where a method based on Green functions and Born approximation has been proposed, to study the nonlinear wave mixing of light fields in metal-dielectric nanostructures of arbitrary geometry \cite{andrey}. 

On a seemingly unrelated matter, Feynman path integrals have proven to be a very elegant, and successful instrument, to describe very complicated systems, ranging from quantum field theoretical problems, to many body problems in condensed matter, and financial markets \cite{kleinert}. Firstly introduced in quantum mechanics by Richard Feynman in his PhD thesis \cite{Feynman}, this method is based on the idea, that the evolution of a quantum system can be described as a sum over all the possible paths, that the system can take, in its evolution from the initial to the final state of its dynamics. Despite path integrals are especially useful in quantum field theories (where they can lead to the introduction of Feynman diagrams, and they represent the correct way, for example, to solve the quantisation problems for the electromagnetic field, due to gauge freedom \cite{das, qft, brown, faddeev}), they have been introduced into optics a number of times, as a way to describe the properties of the electromagnetic field in terms of coherent state representation \cite{path1}, investigate parametric amplification \cite{path2} , atom-field interactions beyond the rotating wave approximation \cite{path3}, quantum decoherence and dephasing in nonlinear spectroscopy \cite{path4}, and the study retardation effects and radiative damping \cite{path5}, to name a few. Moreover, path integral methods have been used in optics to describe beam propagation \cite{path6}, optical fiber communications \cite{path7}, nonparaxial optics \cite{path8}, and, to describe the propagation of the electromagnetic field in homogeneous \cite{path9}, and inhomogeneous \cite{path10} media. In recent years, Bechler \cite{Bechler} proposed a path integral approach to describe quantum electrodynamics in linear dispersive media. Despite the big advantage path integrals represent, in describing very complicated systems in an easy, and manageable way, they have not yet been applied, to the best of our knowledge, to study the problem of quantum electrodynamics in nonlinear arbitrary media.

With this in mind, in this work we then develop a path integral-based theory describing the interactions of the quantised (non-relativistic) electromagnetic field in an arbitrary, nonlinear medium. We develop a complete theory for the quantised electromagnetic field solely, and show, that in this regime,  nonlinear effects can be described by an effective vector quantum field, freely propagating in a dressed vacuum, which accounts for the properties of the medium, which, in our analysis, have not been quantised. We leave the full interaction with the quantised electromagnetic field with quantised matter (i.e., photon-polariton interactions) for future works. Here, in fact, we focus our attention only on optical nonlinearities, with particular emphasis on  second-order processes, and, in particular, to SPDC. Our findings show, how the cross section for second-order processes can be expressed, within the undepleted pump approximation,  in terms of the biphoton propagator, which describes the free propagation of signal and idler modes in the dressed vacuum of our theory, in the presence of nonlinear interactions. Finally, we will present some examples of application of our formalism to simple examples, to show, how our formulas reduce to well known results of nonlinear optics. Moreover, we will also apply our formalism to discuss the link between successive cascaded SPDC, and the onset of squeezing, and we will show, how the nonlinear properties of the medium influence the squeezing parameter.

This work is organised as follows: in Sect. II, we introduce the formalism of path integrals, which constitutes the main tool used throughout the paper, and we calculate the effective action for an electromagnetic field in a linear, arbitrary medium, in terms of the vector potential solely. The quantisation of this effective theory is presented in Sect. III, where also a Fourier representation of the dressed photon propagator is given. In Sect. IV, we discuss the nonlinear interaction of the effective electromagnetic field, while its representation in terms of Feynman diagrams, for the specific case of  $\chi^2$ nonlinearity, is given in Sect. V. In Sect. VI, we apply our formalism to some explicit cases, namely SPDC from a 1D medium, and the generation of squeezed light by repeated cascaded $\chi^{(2)}$-processes. Finally,  a summary of our findings is then given in Sect. VII, and Conclusions are then drawn in Sect. VIII.
\section{Path integral description of the electromagnetic field in arbitrary media}
In this section, we briefly review the basic formalism developed in Ref. \cite{Bechler}, to describe, using the method of path integrals, the propagation of an electromagnetic field in an arbitrary \emph{linear} medium. The basic idea behind the work of Bechler \cite{Bechler}, is to develop an effective theory of the electromagnetic field in an arbitrary medium from a microscopic point of view, where the only accessible degrees of freedom are the electromagnetic ones, and all the information about the medium and its dynamics, can be collectively represented by an effective (i.e., mean field, macroscopic) dielectric constant. This leads to the definition of an effective partition function, i.e., 
\beq\label{eq1}
\mathcal{Z}_{eff}[\vett{E},\vett{B}]=\int\,\mathcal{D}\{\vett{q}\}e^{\frac{i}{\hbar}S[\vett{E},\vett{B};\{\vett{q}\}]}\equiv e^{\frac{i}{\hbar}S_{eff}[\vett{E},\vett{B}]},
\eeq
which constitutes the main quantitiy of interest of this analysis. In the equation above, $\vett{E}\equiv\vett{E}(\vett{x},t)$ ($\vett{B}\equiv\vett{B}(\vett{x},t)$) is the electric (magnetic) field, while $\{\vett{q}\}$ represents a collection of degrees of freedom associated with matter only (such as, as it will be specified later, matter polarisation and loss channels, for example). This effective partition function is obtained from the total action of the system
\beq\label{eq2}
S[\vett{E},\vett{B},\{\vett{q}\}]=\int\,dt\,d^3x\,\mathcal{L}[\vett{E},\vett{B},\{\vett{q}\}],
\eeq
with $\mathcal{L}$ being the Lagrangian density, by integrating on all the possible configurations of the matter degrees of freedom, collectively described by $\{\vett{q}\}$. From a quantum mechanical point of view, this corresponds in tracing out those degrees of freedom, upon which we have either no control, or upon which we have no interest \cite{quantumOpenSystems}. By doing this, in fact, one can describe the dynamics of the electromagnetic field in an arbitrary medium, as the one of a free electromagnetic field``dressed" by the medium itself. Thanks to the integration over the undesired degrees of freedom (pertaining to the medium), in fact, the effect of the medium can be treated macroscopically, by means of an effective dielectric constant \cite{Bechler}.

The knowledge of $\mathcal{Z}_{eff}$ or, analogously, its quantum counterpart, allows us to express any observable quantity associated to the electromagnetic field in terms of ensemble average (in a similar manner, to what it is commonly done in statistical mechanics \cite{landau}) with respect to the partition function $\mathcal{Z}_{eff}$, i.e.,
\beq
\langle f[\vett{E},\vett{B}]\rangle\propto\int\,d\vett{E}\,d\vett{B}\,f[\vett{E},\vett{B}]\mathcal{Z}_{eff}[\vett{E},\vett{B}],
\eeq
or, for the quantum case
\beq
\langle\mathcal{\hat{O}}\rangle\propto\int\,d\vett{E}\,d\vett{B}\,\mathcal{\hat{O}}\mathcal{Z}[\vett{E},\vett{B}].
\eeq
Notice, that in both cases, integration over the electric and magnetic fields is understood as path integration.

It is then clear, that in order to have a viable expression for the partition function $\mathcal{Z}_{eff}$, we need to find a suitable expression for the action $S[\vett{E},\vett{B},\{\vett{q}\}]$ of the particular system considered. This is then the aim of this section. In particular, starting from the Huttner and Barnett action \cite{Huttner}, we will derive the effective action $S_{eff}[\vett{E},\vett{B}]$, and, thus, the classical partition function $\mathcal{Z}_{eff}$. The determination of its quantum counterpart will be the subject of the next sections.

Rather than considering the problem in terms of electric and magnetic fields, as it is done in Ref. \cite{Bechler}, however, here we will derive the effective action in terms of the electromagnetic potentials $\vett{A}\equiv\vett{A}(\vett{x},t)$, and $\Phi\equiv(\vett{x},t)$, which are related to the electric and magnetic fields via the well known relations \cite{jackson}
\bseq
\begin{align}
\vett{E}&=-\frac{\partial\vett{A}}{\partial t}-\nabla\Phi,\\
\vett{B}&=\nabla\times\vett{A}.
\end{align}
\eseq
The main advantage of this approach is, that the effective action, and the resulting effective partition function, will be already in a ready-to-be-quantised form, as they will be expressed in terms of the electromagnetic potentials. The price, that we have to pay for using this representation, however, is given by the fact, that while the electric and magnetic fields are manifestly gauge invariant, the electromagnetic potentials are not, and we therefore need to specify a gauge. This, in particular, is a necessary step for quantisation. 

Although there is a very well-known method for including the gauge choice in the path integral quantisation of the electromagnetic field, i.e., the so-called Faddeev-Popov gauge quantisation \cite{faddeev}, here we assume, for the sake of simplicity, and because we are dealing with non-relativistic fields, to work in a specific gauge, namely the Weyl (temporal) gauge where $\Phi=0$ \cite{WeylGauge}. Although this is a rather exotic gauge for quantum field theory, it is a rather natural choice in this context. From the point of view of optics, in fact, the Weyl gauge corresponds to the situation, in which the vector potential is not transverse anymore, i.e.,  $\nabla\cdot\vett{A}\neq 0$, which is a direct consequence of the fact, that in general, in a dielectric medium, Gauss' law is formulated for the displacement vector $\vett{D}$, rather than for the electric field \vett{E}. Moreover, the Weyl gauge naturally emerges when quantising the electromagnetic field in a dielectric medium, by using the well known method of the noise currents \cite{vogel}.

With this in mind, and following Ref. \cite{Bechler}, we model our system using the Huttner and Barnett Lagrangian density fucntional \cite{Huttner}, which, as a function of the vector potential, can be written as follows:
\barr\label{eq3}
\mathcal{L}[\vett{A},\vett{P},\vett{Y}_{\omega}]&=&\mathcal{L}_{em}[\vett{A}]+\mathcal{L}_{mat}[\vett{P}]+\mathcal{L}_{res}[\vett{Y}_{\omega}]\nonumber\\
&+&\mathcal{L}_{fm}[\vett{A},\vett{P}]+\mathcal{L}_{mr}[\vett{P},\vett{Y}_{\omega}],
\earr
where the first three terms are, respectively, the Lagrangian density of the free electromagnetic field, the matter polarisation field $\vett{P}\equiv\vett{P}(\vett{x},t)$, and the reservoir field $\vett{Y}_{\omega}\equiv\vett{Y}_{\omega}(\vett{x},t)$. Their explicit expressions are reported in Appendix A. The interaction of light with matter is described by the term $\mathcal{L}_{fm}[\vett{A},\vett{P}]$, and it is assumed to be in a minimal coupling form (namely, electric dipole approximiation), i.e.,
\beq\label{eq4}
\mathcal{L}_{fm}[\vett{A},\vett{P}]=g(\vett{x})\,\dot{\vett{A}}\cdot\vett{P},
\eeq
where $g(\vett{x})$ accounts for the medium geometry, and it is a function, which is equal to the dielectric constant of the medium in the region of space filled with the medium, and zero elsewhere. In this model, moreover, the electromagnetic losses are modelled as a reservoir of continuously distributed harmonic oscillators, each characterised by a frequency $\omega$, which interacts only with the matter polarisation field via the term
\beq\label{eq5}
\mathcal{L}_{mr}[\vett{P},\vett{Y}_{\omega}]=-g(\vett{x})\int_0^{\infty}\,d\omega\,f(\omega,\vett{x})\,\vett{P}\cdot\vett{Y}_{\omega},
\eeq
with $f(\omega,\vett{x})$ being the spectral coupling function between the reservoir field, and the matter field \cite{note}.

We can use the Huttner and Barnett Lagrangian density \eqref{eq3} to construct the action $S[\vett{A},\vett{P},\vett{Y}_{\omega}]=\int\,d^3x\,dt\,\mathcal{L}[\vett{A},\vett{P},\vett{Y}_{\omega}]$ of the electromagnetic field propagating in an arbitrary mediuim. Then, we can use Eq. \eqref{eq2} with $\{\vett{q}\}=\{\vett{P},\vett{Y}_{\omega}\}$ to obtain the effective action, and the effective partition function, i.e.,
\beq\label{eq6}
\mathcal{Z}_{eff}[\vett{A}]=\int\,\mathcal{D}\vett{P}\mathcal{D}\vett{Y}_{\omega}e^{\frac{i}{\hbar}S[\vett{A},\vett{P},\vett{Y}_{\omega}]}=e^{\frac{i}{\hbar}S_{eff}[\vett{A}]},
\eeq
where the integration over $\vett{P}$ and $\vett{Y}_{\omega}$ is to be understood as a path integration, and $\mathcal{D}\vett{P}$, and $\mathcal{D}\vett{Y}_{\omega}$ are some suitable positive measures defined on an appropriate manifold, which makes the path integration being correctly defined \cite{kleinert, manifold}. This, ultimately, will allow us to describe the dynamics of the electromagnetic field in such a system, as if it would be a free field ``dressed" by the presence of the dielectric. The details about the calculation of the integrals above are sketched in Appendix B. A more detailed discussion on the general method to approach such integrals   can be found in Ref. \cite{kleinert}, and, for the specific problem at hand, a rather detailed discussion is given in Appendix A of Ref. \cite{Bechler}. 

Using the results highlighted in Appendix B, the effective action for the electromagnetic field propagating in an arbitrary medium assumes the following form
\barr\label{effectiveAction}
&&S_{eff}[\vett{A}]=S_{em}[\vett{A}]\nonumber\\
&+&\frac{1}{2}\int\,dt\,dt'\,d^3x\,g(\vett{x})\dot{\vett{A}}(t,\vett{x})\Gamma(t-t',\vett{x})\dot{\vett{A}}(t',\vett{x}),
\earr
where the expression for the function $\Gamma(t-t',\vett{x})$ is given in Appendix B. Notice, that the above expression of the action is quadratic in the vector potential $\vett{A}$. In field theories, quadratic actions correspond to free fields. In this case, then, we can interpret the effective action $S_{eff}[\vett{A}]$, as the action describing an effective free electromagnetic field, dressed by the presence of the medium. The informations about the medium properties are contained in the function $\Gamma(t-t',\vett{x})$, whose Fourier transform is closely related to the dielectric function of the medium (see, for example, the short discussion in Appendix B, or Ref. \cite{Bechler}).

\section{Quantisation and the effective free theory}\label{section3}
The results obtained above, namely the effective action $S_{eff}[\vett{A}]$,  can be used to predict the classical dynamics of the electromagnetic field in an arbitrary medium. If one is interested in its quantum properties, however, a further step is needed, to quantise the theory represented by $\mathcal{Z}_{eff}$. In quantum field theory, this is typically done by using the Faddeev-Popov gauge quantisation method \cite{faddeev, qft}, which allows, once the electromagnetic field has been represented in the form of a path integral, to automatically take into account the gauge dependence of the 4-potential, thus allowing a correct quantisation of the electromagnetic field. This is typically obtained by first inserting a coupling term in the action of the electromagnetic field, which takes into account the interaction of  the electromagnetic field with a fictitious source current $\vett{J}(t,\vett{x})$, i.e.,
\beq\label{quantumAction}
S_q[\vett{A},\vett{J}]=S_{eff}[\vett{A}]+\int\,dt\,d^3x\,\vett{J}\cdot\vett{A}
\eeq
and then integrate over all the possible field configurations $\vett{A}$, with a suitable measure, which accounts for gauge freedom \cite{qft,nostroarXiv}. For the case under analysis, however, since we are dealing with non relativistic fields, and since we have already made a choice of gauge (i.e., the Weyl gauge) to obtain the classical partition function $\mathcal{Z}_{eff}$, the vector potential is already uniquely determined by $\vett{A}$, and path integration now gives a properly regularised result \cite{note3}.

The quantity of interest for our calculations is therefore the free quantum partition function
\beq
\mathcal{Z}_0[\vett{J}]=\int\,\mathcal{D}\vett{A}\,e^{\frac{i}{\hbar}S_q[\vett{A},\vett{J}]},
\eeq
which is used to calculate the dynamical properties of the electromagnetic field. To compute the above integral, we use the method highlighted in Appendix B. To do that, we first need to write $S_q[\vett{A},\vett{J}]$ as a quadratic form of the type $(\vett{A},\hat{R}\vett{A})+(\vett{b},\vett{A})$, where $\vett{b}=\vett{J}\delta(\vett{x}-\vett{x}')$, and the explicit expression of the operator $\hat{R}$ is given in Appendix C. If we now define the dressed photon propagator  as $D(t-t',\vett{x}-\vett{x}')\equiv\hat{R}^{-1}(t-t',\vett{x}-\vett{x}')$, and we solve the above integral using Gaussian integration, we get the following result:
\beq\label{freeTheory1}
\mathcal{Z}_0[\vett{J}]=\mathcal{N}_0e^{\frac{i}{2\hbar}\int\,dt\,dt'\,d^3x\,d^3x'\,\vett{J}(t,\vett{x})D(t-t',\vett{x}-\vett{x}')\vett{J}(t',\vett{x'})},
\eeq
where $\mathcal{N}_0$ is a suitable normalisation constant, which is related to $\mathcal{Z}_0[0]$. Notice, moreover, that the dressed photon propagator $D(t-t',\vett{x}-\vett{x'})$ is a tensor of rank two, as it connects different components of the vector fields $\vett{J}(t,\vett{x})$, and $\vett{J}(t',\vett{x}')$.  A closer inspection to the above equation reveals, that the action
\beq\label{freeAction}
S_0[\vett{J}]=\frac{1}{2}\int\,dt\,dt'\,d^3x\,d^3x'\,\vett{J}(t,\vett{x})D(t-t',\vett{x}-\vett{x}')\vett{J}(t',\vett{x'}),
\eeq
appearing in the exponent of Eq. \eqref{freeTheory1}, has the typical form of the action of a free vector field $\vett{J}$, whose dynamics are described by the dressed propagator $D_{\mu\nu}(t-t',\vett{x}-\vett{x'})$. Notice, moreover, that the knowledge of the quantum partition function $\mathcal{Z}_0[\vett{J}]$ is sufficient, to derive all the dynamical properties of the field. In particular, for example, the propagator can be obtained through functional derivation of $\mathcal{Z}_0[\vett{J}]$ with respect to the fields $\vett{J}$, i.e. \cite{qft},
\beq\label{eqProp}
D_{\mu\nu}(t-t',\vett{x}-\vett{x'})=i\hbar\frac{\delta\mathcal{Z}_0[\vett{J}]}{\delta J_{\mu}(t,\vett{x})\delta J_{\nu}(t',\vett{x}')}\Bigg|_{\vett{J}=0},
\eeq
where $\delta/\delta J$ is the functional derivative \cite{kleinert}, defined such that
\beq
\frac{\delta J_{\mu}(t,\vett{x})}{\delta J_{\nu}(\tau,\vett{y})}=\delta_{\mu\nu}\delta(\vett{x}-\vett{y})\delta(t-\tau).
\eeq
Equation \eqref{eqProp} links  the $\mu$-component of the field at time $t$ and position $\vett{x}$, with the $\nu$-component of the field, at time $t'$ and position $\vett{x}'$, through the dressed photon propagator $D_{\mu\nu}(t-t',\vett{x}-\vett{x}')$. It is not difficult to show, that, according to standard path integral theory \cite{das}, the dressed photon propagator defined above can be put in relation to the field operators as follows
\beq\label{eq15}
D_{\mu\nu}(t-t',\vett{x}-\vett{x}')=\frac{i}{\hbar}\left\langle\mathcal{T}\left\{\hat{A}_{\mu}(t,\vett{x})\hat{A}_{\nu}(t',\vett{x}')\right\}\right\rangle,
\eeq
where $\mathcal{T}\{.\}$ represents the time-ordering operator, and $\langle\hat{O}\rangle$ is the expectation value of the operator $\hat{O}$ \cite{vogel}.

From the point of view of optics, Eq. \eqref{eqProp} tells us, how the $\mu$-component of the electromagnetic field influences its $\nu$-component, i.e., their correlation, while propagating in the medium. Such behaviour, for example, is typical in anisotropic crystals, where, in general, each field component impinging on the crystal, affects the propagation of every other one \cite{boyd}.
\subsection{Fourier representation of the dressed photon propagator}
The result given by Eq. \eqref{eqProp} is very general, and, in principle, allows us to describe the propagation of the electromagnetic field in an arbitrary medium, in time domain, provided, that we know the temporal properties of the medium. In optics, however, the information on the dielectric function of a certain medium is typically given in frequency domain, as it is easier, and more practical from an experimental point of view, to access informations as a function of the frequency \cite{jackson}. Moreover, most of the nonlinear optical phenomena of interest, such as SPDC, and parametric processes,  are typically described in frequency domain \cite{boyd}. For this reason, it is useful to find a suitable representation for the dressed photon propagator in frequency domain, rather than in time domain.  Developing a frequency-domain path integral, therefore, will allow us to directly calculate quantities, such as the photon generation probability, or the efficiency of certain nonlinear processes, in a form, that can be directly compared with the correspondent quantities available in literature. 

To this aim, then, let us notice, that the dressed photon propagator $D(t-t',\vett{x}-\vett{x}')$ is the Green function of the integro-differential operator $\hat{R}$, defined in Appendix C, i.e.,
\barr
&&\Big[\Big(\varepsilon_0\frac{\partial^2}{\partial t^2}-\frac{1}{\mu_0}\nabla^2\Big)\delta_{\mu\alpha}+\frac{1}{\mu_0}\frac{\partial^2}{\partial x_{\mu}\partial x_{\alpha}}\Big]D_{\alpha\nu}(x-x')\nonumber\\
&+&g(\vett{x})\int\,d\tau\,\Big[\frac{\partial^2}{\partial t^2}\Gamma(t-\tau,\vett{x})\Big]D_{\mu\nu}(\tau-t',\vett{x}-\vett{x}')\nonumber\\
&=&\delta_{\mu\nu}\delta(x-x'),
\earr
where summation over repeated indices is implicitly understood, and the shorthand $x\equiv\{t,\vett{x}\}$ has been used. If we take the Fourier transform (with respect to time $t$) of the above equation, call $G_{\mu\nu}(\omega,\vett{x})$ the Fourier transform of the dressed propagator, consider only positive frequencies (namely, we restrict our analysis to $\omega>0$), and use the results of  Appendix B to link the Fourier transform of $\Gamma(t-t',\vett{x})$ to the dielectric function of the medium $\varepsilon(\omega,\vett{x})$, we can rewrite the above equation in the following way:
\begin{widetext}
\beq\label{helmholtz}
\Big[\Big(-\delta_{\mu\alpha}\nabla^2+\frac{\partial^2}{\partial x_{\mu}\partial x_{\alpha}}\Big)-\frac{\omega^2}{c^2}\varepsilon(\omega,\vett{x})\delta_{\mu\alpha}\Big]G_{\alpha\nu}(\omega, \vett{x}-\vett{x}')=\mu_0\delta_{\mu\nu}\delta(\vett{x}-\vett{x}').
\eeq
\end{widetext}
This allows us to interpret $G_{\mu\nu}(x)$ as the Green function of the Helmholtz equation for a monochromatic electromagnetic field, propagating in an arbitrary medium, whose properties are described by the dielectric function $\varepsilon(\omega,\vett{x})$ \cite{vogel}. This result also gives us the possibility to consider the effects and properties of the medium from a macroscopic point of view only, through the dielectric function $\varepsilon(\omega,\vett{x})$, which can be calculated, and experimentally measured, using different techniques \cite{epsilon1,epsilon2,epsilon3}. 

We can then rewrite Eq. \eqref{freeTheory1}, using the Fourier representation $G_{\mu\nu}(\omega,\vett{x})$ of the photon propagator, given by the equation above, as follows:
\beq\label{eq18}
\mathcal{Z}_0[\vett{J}]=e^{\frac{i}{2\hbar}\int\,d\omega\,d^3x\,d^3x'\,J_{\mu}(\omega,\vett{x})G_{\mu\nu}(\omega,\vett{x}-\vett{x}')J_{\nu}(\omega,\vett{x}')}.
\eeq
This is the first result of our work. The dynamics of the electromagnetic field in an arbitrarily shaped, linear medium, can be interpreted as the ones of an effectively free field, propagating in a ``vacuum" dressed by the properties of the medium, which define the photon propagator $G_{\mu\nu}(\omega,\vett{x})$.
\section{interacting theory}
The partition function given in Eq. \eqref{eq18} fully describes the dynamics and correlation properties of photons in an arbitrary, linear medium. Once the dressed photon propagator is known, in fact, the standard methods of quantum field theory can be applied to Eq. \eqref{eq18} to extract any quantity of interest, such as correlations, or absorption/emission rates \cite{das}. However, Eq. \eqref{eq18} cannot describe any nonlinear interaction of the field in the medium. To do that, we need to introduce a proper interaction Lagrangian, which correctly takes into account for the nonlinear interaction of the electromagnetic field, for the various relevant kinds of optical nonlinearities \cite{boyd}.

To implement such Lagrangian, we first recall, that optical nonlinearities are typically quite small in magnitude, and can be therefore treated within the framework of perturbation theory. This will allow us to express the interaction Lagrangian in a suitable power series expansion, which will facilitate the introduction of Feynman diagram, and the calculation of the cross sections of the various nonlinear processes. 

Then, we can construct a suitable interaction Lagrangian, by looking at the way optical nonlinearities are inserted in optics.  In such cases, in fact, nonlinearities enter Maxwell's equation through the field polarisation $\boldsymbol\Pi$ , which is typically expressed  in power series of the electric field, i.e., $\boldsymbol\Pi=\varepsilon_0\left(\chi^{(1)}\vett{E}+\chi^{(2)}\vett{E}^2+\chi^{(3)}\vett{E}^3+\cdots\right)$, where $\chi^{(n)}\equiv\chi^{(n)}(\omega,\vett{x})$ is the $n$-th order susceptibility tensor, with $1+\chi^{(1)}=\varepsilon$ being the dielectric function of the medium \cite{boyd}. It is not difficult to show, that this kind of interaction can be generated by a Lagrangian of the form
\barr\label{intLag}
&&\mathcal{L}_{int}[\vett{A}]=\sum_{n=2}^{\infty}\frac{(i\omega)^{n+1}}{(n+1)!}\boldsymbol\chi^{(n)}\cdot\vett{A}^{n+1}\nonumber\\
&=&-\frac{i\omega^3}{3!}\chi_{\mu\nu\sigma}^{(2)}A_{\mu}A_{\nu}A_{\sigma}+\frac{\omega^4}{4!}\chi_{\mu\nu\sigma\tau}^{(3)}A_{\mu}A_{\nu}A_{\sigma}A_{\tau}\nonumber\\
&+&\text{higher orders},
\earr
where summation over repeated indices has been implicitly understood, and $\{\mu,\nu,\sigma,\tau\}\in\{x,y,z\}$. With this result, we can write the partition function for the interacting, nonlinear, theory as follows \cite{qft}
\beq\label{intZ}
\mathcal{Z}[\vett{J}]=\mathcal{N}e^{\frac{i}{\hbar}\int\,d\omega\,d^3x\,\mathcal{L}_{int}\left[\frac{1}{i}\frac{\delta}{\delta\vett{J}}\right]}\mathcal{Z}_0[\vett{J}],
\eeq
where $\mathcal{N}$ is a suitable normalisation constant, and the argument of the interaction Lagrangian appearing in the exponent above has the meaning of replacing every entry of the vector potential $\vett{A}$ with a functional derivative with respect to the correspondent current component, according to the identity \cite{qft}
\beq
\langle A_{\mu}(\omega,\vett{x})\rangle=\frac{1}{i}\frac{\delta\mathcal{Z}_0[\vett{J}]}{\delta J_{\mu}(\omega,\vett{x})}\Bigg|_{\vett{J}=0}.
\eeq
Equation \eqref{intZ} can be then expanded perturbatively, by assuming that the interaction Lagrangian is defined in terms of a small parameter, which in the present case, is the strength of the nonlinear interaction. In particular, we have to assume, that the magnitude of each nonlinear susceptibility tensor appearing in Eq. \eqref{intLag} is very small (compared to the linear susceptibility), i.e., $|\chi^{(n)}|\ll\chi^{(1)}$, $\forall n\geq 2$, and that the higher order nonlinearities are progressively smaller, i.e., $|\chi^{(n+1)}|\ll |\chi^{(n)}|$, $\forall n\geq 2$. Under these assumptions, which are verified for typical nonlinear optical materials, we can expand the exponential term appearing in Eq. \eqref{intZ} into a power series, to obtain the following result
\barr\label{intFull}
\mathcal{Z}[\vett{J}]&=&\mathcal{Z}_0[\vett{J}]+\sum_{k=2}^{\infty}\mathcal{Z}^{(k)}[\vett{J}]+\mathcal{Z}_{cross}[\vett{J}],
\earr
where $\mathcal{Z}_0[\vett{J}]$ is the partition function of the free theory, as given by Eq. \eqref{eq18}, $\mathcal{Z}^{(k)}[\vett{J}]$ represents the correction to the partition function due to the presence of $k$-th order nonlinearity in the medium, whose explicit form is given by
\barr
&&\mathcal{Z}^{(k)}[\vett{J}]=\sum_{n=1}^{\infty}\frac{i^{k+1}}{n!(k+1)!}\left(\frac{i}{\hbar}\right)^n\nonumber\\
&\times&\left[\int\,d\omega\,d^3x\,\omega^{k+1}\,\boldsymbol\chi^{(k)}\cdot\left(\frac{1}{i}\frac{\delta}{\delta\vett{J}}\right)^{k+1}\right]^n\mathcal{Z}_0[\vett{J}],
\earr
and $\mathcal{Z}_{cross}[\vett{J}]$ is the cross nonlinearity term,  which contains informations about the interplay between the different orders of nonlinearities (i.e., it contains terms proportional to $\Pi_k\boldsymbol\chi^{(k)}\boldsymbol\chi^{(k+1)}\boldsymbol\chi^{(k+2)}\cdots$). 

In practical situations, however, this term can be neglected, as it is typically of higher order, with respect to the order of the considered nonlinearity. Let us assume, for example, to consider a medium, in which both second-, and third-order nonlinearities are present. Typical values of the magnitudes of such nonlinearities are, respectively $|\chi^{(2)}|\simeq 10^{-12}$ $m/V$, and  $|\chi^{(3)}|\simeq 10^{-24}$ $m^2/V^2$ \cite{boyd}.  A cross-interaction term involving both nonlinear processes, then, will be of the order of $|\chi^{(2)}|\times|\chi^{(3)}|\simeq 10^{-36}$ $m^3/V^3$, which is orders of magnitude smaller, that the two separate processes, and can therefore be neglected. 

Accounting for this approximation, we then have
\barr\label{intFinal}
&&\mathcal{Z}[\vett{J}]=\Bigg\{1+\sum_{k=2}^{\infty}\sum_{n=1}^{\infty}\frac{i^{k+1}}{n!(k+1)!}\left(\frac{i}{\hbar}\right)^n\nonumber\\
&\times&\left[\int\,d\omega\,d^3x\,\omega^{k+1}\,\boldsymbol\chi^{(k)}\cdot\left(\frac{1}{i}\frac{\delta}{\delta\vett{J}}\right)^{k+1}\right]^n\Bigg\}\mathcal{Z}_0[\vett{J}],
\earr
where the index $k$ runs through the nonlinearities, that are present in the medium, and the index $n$ accounts for the various terms of the power series expansion of the exponential term in Eq. \eqref{intZ}. This result represents the most general nonlinear interaction of the electromagnetic field, with a medium containing all orders of optical nonlinearities, each one described by its own nonlinear susceptibility  $\boldsymbol\chi^{(n)}$. Written in this form, moreover, the above expression can be easily translated in the language of Feynman diagrams.

 In optics, however, typically only second- and third-order processes are considered, as they are, wuth the present technology, the only accessible nonlinear processes experimentally. For this reason, then, it makes sense to specialise this formalism, to the study of $\chi^{(2)}-$, and $\chi^{(3)}-$processes, solely. In this work, however, we limit our attention to $\chi^{(2)}-$processes, leaving the more rich structure of $\chi^{(3)}-$nonlinearities to future investigations.
\subsection{Interaction Lagrangian for second order nonlinear processes}
Second-order nonlinear processes involve the interaction of three photons, often called pump, signal and idler modes \cite{boyd}. In this case, the nonlinear susceptibility is a rank 3 tensor $\chi_{\sigma\mu\nu}(\omega,\vett{x})$, and, according to Eq. \eqref{intLag}, the interaction Lagrangian density describing such processes can be written as 
\beq\label{lag2}
\mathcal{L}_{int}^{(2)}[\vett{A}]=\frac{1}{3!}\chi_{\sigma\mu\nu}(\omega,\vett{x})A_{\sigma}^{(p)}A_{\mu}^{(s)}A_{\nu}^{(i)},
\eeq
where the superscripts $\{p,s,i\}$ stands for pump, signal, and idler photon modes, respectively. Notice moreover, that in the equation above, the nonlinear susceptibility $\chi_{\sigma\mu\nu}(\omega,\vett{x})$ has been redefined in such a way, to include the term $-i\omega^3$ appearing in Eq. \eqref{intLag}, for later convenience.

With the above interaction Lagrangian, it is possible to describe all possible second-order processes, such as second harmonic generation (SHG), sum (difference) frequency generation (SFG/DFG), and (spontaneous) parametric down conversion (SPDC). While SHG consist in the conversion of two degenerate signal and idler modes into a pump one (with SPDC being, practically, its inverse process), SFG (DFG) describes the scattering of a signal (idler) photon, into an idler (signal) one, mediated by the presence of a pump photon \cite{boyd,loudon}. SPDC, in particular, is of great importance in modern optical laboratories, as it is the main mechanisms used to generate entangled photon pairs \cite{quantumNLO} and squeezed light \cite{loudon}.
\subsection{Undepleted pump approximation and the quantum optical dressed vacuum}
The interaction Lagrangian described by Eq. \eqref{lag2} is exact, and correctly describes all the possible second-order nonlinear processes. However, it implicitly assumes, that all three modes involved in the process, namely signal, idler, and pump, are quantum fields, and, ideally, contain only few photons. In practical situations, however, nonlinear optics experiments are typically carried out by using a very intense pump, which stimulates the onset of nonlinear processes. The reason behind this is very simple: as nonlinear processes are very weak, high intensities , i.e., a high number of photons, are needed, in order to make the process probable enough to be observed. Under these working condition, then, the so-called undepleted pump approximation is used. Under this assumption, the pump mode $A_{\sigma}^{(p)}$ contains a large number of photons, and it is often described in terms of coherent states. The occasional conversion of energy from the pump to the signal and idler modes (regulated by the energy conservation constraint $\omega_p=\omega_s+\omega_i$), then, does not affect the number of photons contained in the pump mode  (i.e., the intensity of the pump beam), which, in first approximation, can be considered to remain constant. For this reason, therefore, the pump mode is often considered as a classical object, and it enters in the dynamics only parametrically, de facto contributing in the definition of an effective nonlinear coefficient. 

In our theory, we can account for this approximation, by promoting the pump field $A_{\sigma}^{(p)}$ to be a classical field, and by introducing the effective nonlinear coupling constant  $\lambda_{\mu\nu}(\omega \vett{x})=\chi_{\sigma\mu\nu}(\omega,\vett{x})A_{\sigma}^{(p)}$, so that Eq. \eqref{lag2} can be written as
\beq\label{effIntLag}
\mathcal{L}_{int}^{(2)}[\vett{A}]=\frac{1}{3!}\lambda_{\mu\nu}(\omega,\vett{x})A_{\mu}^{(s)}A_{\nu}^{(i)}.
\eeq
Notice, how applying the undepleted pump approximation to the interaction Lagrangian \eqref{lag2} resulted in removing the pump mode from the dynamics, thus leaving an effective two-photon interaction term, i.e., a quadratic effective interaction Lagrangian.

Before proceeding any further, it is worth to discuss the consequences of this result. First, notice, that quadratic Lagrangians in quantum field theories typically describe the dynamics of free fields. In this case, the quadratic nature of the interaction Lagrangian comes from a form of effective self-interaction of the field. This, ultimately, hints at the possibility of describing the nonlinear interaction of the electromagnetic field in an arbitrary medium, in terms of the dynamics of a free, self-interacting field, i.e., a non-abelian gauge field \cite{gauge1}.

Second, and more relevant to the present discussion, is the meaning of the effective coupling constant  $\lambda_{\mu\nu}(\omega,\vett{x})$, defined in terms of the pump field  $A_{\sigma}^{(p)}$. We have seen in the previous section, in fact, that the propagation of an electromagnetic field in an arbitrary medium, can be understood in terms of the partition function $\mathcal{Z}_0[\vett{J}]$, describing the dynamics of a free effective electromagnetic field, dressed by the medium.

In other words, the vacuum state of such a free theory automatically accounts for the presence of the medium, as it is proven, by the fact, that the vacuum expectation value of the two-point correlation function of such an effective electromagnetic field gives as result the dressed photon propagator $G_{\mu\nu}(\omega,\vett{x})$ [see Eq. \eqref{eq15}]. The introduction of the nonlinear interaction, and the correspondent undepleted pump approximation, on the other hand, adds another layer of structure to this vacuum, practically dressing it with the properties of the pump beam. The true vacuum state of the effective electromagnetic field, then, can be written as
\beq
\ket{0}\equiv\ket{\{0_{\omega}\}_D;\omega_p},
\eeq
where $\{0_{\omega}\}_D$ is a shorthand for describing all the frequency modes of the dressed electromagnetic field, and $\omega_p$ highlights the fact, that the vacuum state is dressed by the pump mode, which, under the undepleted pump approximation, parametrises the nonlinear interaction. Within this framework, for exmaple, SPDC can be then described as the spontaneous generation of a signal-idler photon pair from the vacuum, i.e., $\ket{0}\rightarrow\ket{1_s,1_i}=\ket{\{0_{\omega},\cdots1_{\omega_s}\,,1_{\omega_i},\,\cdots\,0_{\omega},\cdots\};\omega_p}$, where $1_{\omega_{s,i}}$ indicates, that a signal (idler) photon has been generated in the mode at frequency $\omega_{s,i}$, respectively, according to the energy conservation contraint $\omega_s+\omega_i=\omega_p$. 

This is the second result of our work. Within the undepleted pump approximation, the nonlinear dynamics of the electromagnetic field in a $\chi^{(2)}$-medium, can be understood as the free dynamics of a self-interacting field, whose vacuum has been suitably dressed, accounting for both the properties of the medium the electromagnetic field is propagating into, and the properties of the pump beam, used to trigger the nonlinear processes.

\subsection{Partition function for second order nonlinear phenomena in the undepleted pump approximation}
We are now in the position to calculate the explicit expression of the partition function $\mathcal{Z}[\vett{J}]$, for $\chi^{(2)}$-nonlinearities  in the undepleted pump approximation. To do that, we substitute the expression of the interaction Lagrangian density given by Eq. \eqref{effIntLag} into the expression of $\mathcal{Z}_k[\vett{J}]$, with $k=2$, and, for the sake of simplicity, we limit ourselves to consider only the first order of the expansion of the exponential term appearing in Eq. \eqref{intZ}. Higher orders, in fact, can be easily derived using the same line of reasoning presented here. With a bit of algebra, it is not difficult to show, that Eq. \eqref{intFinal} can be then written as
$\mathcal{Z}[\vett{J}]=\mathcal{Z}_0[\vett{J}]+\mathcal{Z}_1[\vett{J}]+ \mathcal{O}(\lambda^2)$, where $\mathcal{Z}_0[\vett{J}]$ is given by Eq. \eqref{eq18}, and
\barr
\mathcal{Z}_1[\vett{J}]&=&\frac{i}{3!\hbar}\int\,d^4z\,\lambda_{\alpha\beta}(z)\frac{\delta^2\mathcal{Z}_0[\vett{J}]}{\delta J_{\alpha}(z)\delta J_{\beta}(z)}\nonumber\\
&=&\Bigg\{-\frac{i}{2\hbar}\int\,d^4z\,\lambda_{\alpha\beta}(z)G_{\alpha\beta}(0)\nonumber\\
&+&\left(\frac{1}{2\hbar}\right)^2\int\,d^4x\,d^4y\,d^4z\,\lambda_{\alpha\beta}(z)J_{\mu}(x)\nonumber\\
&\times&\left[G_{\mu\alpha}(x-z)G_{\beta\nu}(z-y)\right]J_{\nu}(y)\Bigg\}\mathcal{Z}_0[\vett{J}],
\earr
where summation over repeated indices is implicitly understood, and the shorthands $x=\{\omega',\vett{x}'\}$, $y=\{\omega'',\vett{x}''\}$, and $z=\{\omega,\vett{x}\}$ have been used, for the sake of clarity.

Formally, the first term diverges. However, this amounts to a loop correction of the vacuum state (in terms of Feynman diagrams), and can be properly regularised through standard renormalisation \cite{qft}. In terms of nonlinear optics, this term corresponds to the process of a signal-idler photon pair being created from the ``dressed" vacuum, and immediately recombine. Since we are under the undepleted pump approximation, this process is of no interest, since it only amounts to a local fluctuation of the pump intensity, and we can therefore neglect it. If we then normalise the partition function $\mathcal{Z}[\vett{J}]$ to the non-interacting partition function $\mathcal{Z}_0[\vett{J}]$, and consider only the interacting part (the zero-order term would be the identity, which corresponds to the non-interacting case, and thus gives no information on the interaction), we obtain the following result
\barr\label{chi2Z}
\frac{\mathcal{Z}[\vett{J}]}{\mathcal{Z}_0[\vett{J}]}&=&\left(\frac{1}{2\hbar}\right)^2\int d^4xd^4yJ_{\mu}(x)\mathcal{X}^{(2)}_{\mu\nu}(x-y)J_{\nu}(y),
\earr

where
\beq\label{biphotonProp}
\mathcal{X}^{(2)}_{\mu\nu}(x-y)=\int\,d^4z G_{\mu\alpha}(x-z)\lambda_{\alpha\beta}(z)G_{\beta\nu}(z-y),
\eeq
is the two-mode (or bi-photon) propagator, which describes the dynamics of the signal and idler fields under the effect of the nonlinear interaction.

This is the main result of our work. Second-order nonlinear processes in the undepleted pump approximation are described by the quantum partition function given in Eq. \eqref{chi2Z}, which is in the form of the partition function of a free quantum field, characterised by the bi-photon propagator $\mathcal{X}_{\mu\nu}^{(2)}(x-y)$. This result constitutes a generalisation of the traditional bi-photon wave function approach to nonlinear optical processes \cite{biphot1,biphot2,biphot3}, since it does not only contain information about the various frequency modes involved in the dynamics, as in the traditional approach, but it also contains information about the spatial distribution of the electromagnetic field inside the medium, and the properties of the medium itself.
\section{Feynman diagram representation for $\mathcal{Z}[\vett{J}]$}
%
%
%
%
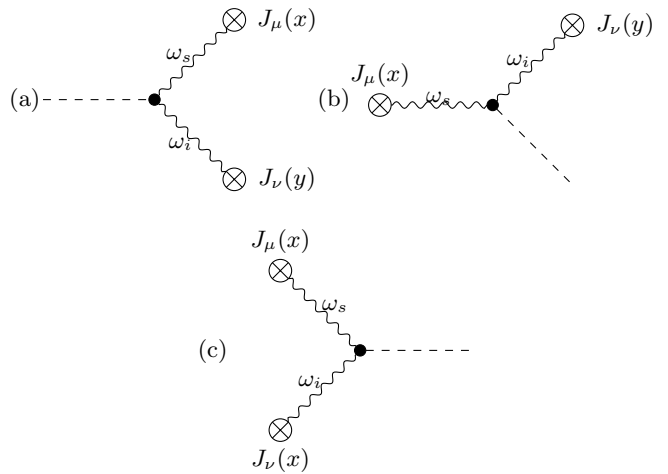
\begin{figure}[!t]
\centering
\begin{tikzpicture}[
 image/.style = {
                 inner sep=0pt, outer sep=0pt},
node distance = 1mm and 1mm
                        ] 
\node [image] (frame1){
(a)\begin{minipage}{0.2\textwidth}
\begin{tikzpicture}
\begin{feynman}
\vertex  (a0){};
\vertex [dot, right= of a0] (a){};
\vertex [crossed dot, above right= of a] (w1){};
\vertex [right = 0.7 cm of w1] (c1) {\( J_{\mu}(x)\)};
\vertex [crossed dot, below right= of a] (w2){};
\vertex [right = 0.7 cm of w2] (c2) {\( J_{\nu}(y)\)};
\diagram* {
(a0) -- [scalar] (a) -- [boson, edge label= \(\omega_s\)] (w1),
(a) -- [boson, edge label' = \(\omega_i\)] (w2),
};
\end{feynman}
\end{tikzpicture}
\end{minipage}
};
\node [image,right=of frame1] (frame2) {
(b)\begin{minipage}{0.2\textwidth}
\begin{tikzpicture}
\begin{feynman}
\vertex [crossed dot] (a0){};
\vertex [above = 0.4 cm of a0] (c1) {\( J_{\mu}(x)\)};
\vertex [dot, right= of a0] (a){};
\vertex [crossed dot, above right= of a] (w1){};
\vertex [below right= of a] (w2){};
\vertex [right = 0.7 cm of w1] (c2) {\( J_{\nu}(y)\)};
\diagram* {
(a0) --  [boson, edge label= \(\omega_s\)]  (a) --[boson, edge label= \(\omega_i\)]  (w1),
(a) --  [scalar]  (w2),
};
\end{feynman}
\end{tikzpicture}
\end{minipage}
};
\hspace{2.5cm}
\node[image,below= 0.5 cm of frame1] (frame3) {
(c)\begin{minipage}{0.2\textwidth}
\begin{tikzpicture}
\begin{feynman}
\vertex [dot] (a0){};
\vertex [crossed dot, above left= of a0] (b) {};
\vertex [crossed dot, below left= of a0] (c) {};
\vertex [right= of a0] (d) {};
\vertex [above=0.4cm of b](c1) {\(J_{\mu}(x)\)};
\vertex [below=0.4cm of c](c2) {\(J_{\nu}(x)\)};
\diagram* {
(a0) -- [boson, edge label'= \(\omega_i\)] (c),
(a0) -- [boson, edge label'= \(\omega_s\)] (b),
(a0) -- [scalar] (d),
};
\end{feynman}
\end{tikzpicture}
\end{minipage}
};
\end{tikzpicture}
\caption{Relevant Feynman diagrams for $\chi^2$-processes, corresponding to the various terms composing $\mathcal{Z}[\vett{J}]$. (a) Generation of a signal ($\omega_s$)-idler ($\omega_i$) photon pair from the dressed vacuum (SPDC). (b) Scattering of a photon from the signal mode $\omega_s$, to the idler mode $\omega_i$ (DFG). (c)  Annihilation of a signal-idler photon pair in the dressed vacuum (SFG). The Feynman diagrams depicted in this figure, and all the other throughout the whole manuscript, have been generated using TikZ-Feynman \cite{tikZfeynman}}
\label{figure1}
\end{figure}
%
%
To make our result more clear, and easily understandable, in this section we rewrite Eq. \eqref{intZ} in terms of Feynman diagrams, and we then introduce the relevant diagrams for $\chi^{(2)}$-nonlinearities, i.e., for Eq. \eqref{chi2Z}. To start with, we make use of the standard methods of quantum field theory, to rewrite Eq. \eqref{intZ}, for the case of second-order nonlinearity, in the following form, which can be then readily translated into the Feynman diagrams formalism:
\barr\label{feynman1}
&&\mathcal{Z}[\vett{J}] =\mathcal{N}e^{\frac{i}{\hbar}\int\,d\omega\,d^3x\,\mathcal{L}_{int}\left[\frac{1}{i}\frac{\delta}{\delta\vett{J}}\right]}\mathcal{Z}_0[\vett{J}]\nonumber\\
&=&\mathcal{N}\sum_{V=0}^{\infty}\frac{1}{V!}\Bigg[\frac{i}{3!\hbar}\int\,d^4z\lambda_{\alpha\beta}(z)\left(\frac{1}{i}\frac{\delta^2}{\delta J_{\alpha}(z)\delta J_{\beta}(z)}\right)\Bigg]^V\nonumber\\
&\times&\sum_{P=0}^{\infty}\frac{1}{P!}\Bigg[\frac{i}{2\hbar}\int\,d^4xd^4y\,J_{\mu}(x)G_{\mu\nu}(x-y)J_{\nu}(y)\Bigg]^P
\earr
The expression above has been obtained fromEq. \eqref{intZ}, with Eq. \eqref{effIntLag}, and by performing a double series expansion, one with respect to the interaction parameter $\lambda_{\alpha\beta}(z)$, and the other, with respect to the current $\vett{J}$. Notice, moreover, that while the index $P$ is connected to the free partition function, and, thus, counts the number of propagators to be displayed in the Feynman diagram (the last line in the above equation, in fact, is the power series expansion of $\mathcal{Z}_0[\vett{J}]$), the index $V$ is related to the nonlinear interaction, and, essentially, counts the number of vertices (i.e., interactions).

If we compare Eqs. \eqref{feynman1}, and Eq. \eqref{chi2Z}, it is not difficult to see, that Eq. \eqref{feynman1} reduces to Eq. \eqref{chi2Z} for $V=1$, and $P=2$. This means, that Eq. \eqref{chi2Z} only contains single interaction events ($V=1$ means, in fact, that only one vertex is allowed in the correspondent Feynman diagram), and two photon modes, represented by 2 propagators (hence, $P=2$).

We can then introduce the Feynman rules for $\chi^{(2)}$-nonlinearities as follows:
\begin{itemize}
\item a dashed line segment represents the dressed vacuum state $\ket{0}\equiv\ket{\{0_{\omega}\}_D;\omega_p}$;
\item a wiggled line represents the dressed photon propagator $G_{\mu\nu}(x-y)/(i\hbar)$;
\item a dashed crossed dot indicates an external source current $J_{\mu}(x)$;
\item a black dot indicates a vertex, where at maximum 2 lines can join (the dashed line representing the dressed vacuum does not count, towards this limit). To each vertex, representing the nonlinear interaction, the term $(i/3!\hbar)\int\,d^4z\lambda_{\alpha\beta}(z)$ is associated;
\item at each vertex, energy conservation must be fulfilled.
\end{itemize}
%
%
In other terms, the following identifications have to be understood:
\barr
\begin{gathered}
\begin{tikzpicture}
\begin{feynman}
\vertex (a) {};
\vertex [right= of a] (b) {};

\diagram * {
(a) -- [scalar] (b)
};
\end{feynman}
\end{tikzpicture}
\end{gathered}
&\rightarrow&\ket{0},\\
\begin{gathered}
\begin{tikzpicture}
\begin{feynman}
\vertex [right= of a] (b) {};
\diagram* {
 (a) -- [boson] (b) 
};
\end{feynman}
\end{tikzpicture}
\end{gathered}
&\rightarrow&\frac{1}{i\hbar}G_{\mu\nu}(x-y),\\
\begin{gathered}
\begin{tikzpicture}
\begin{feynman}
\vertex [crossed dot] (a) {};
\diagram* {
(a0) (a) 
};
\end{feynman}
\end{tikzpicture}
\end{gathered}
&\rightarrow&i\int\,d^4x \,J_{\mu}(x),\\
\begin{gathered}
\begin{tikzpicture}
\begin{feynman}
\vertex [dot] (a){};
\diagram* {
(a0) (a)
};
\end{feynman}
\end{tikzpicture}
\end{gathered}
&\rightarrow&\frac{i}{3!\hbar}\int\,d^4z\,\lambda_{\alpha\beta}(z).
\earr
%
%
Having defined the above rules for drawing Feynman diagrams, gives us the possibility, to calculate the partition function $\mathcal{Z}[\vett{J}]$, and the related correlation functions $\langle A_{\mu}(x_1)\cdots A_{\nu}(x_n)\rangle$ in a very intuitive way. 

Let us illustrate this with an example, which makes use of the Feynman diagram representation of the free partition function $\mathcal{Z}_0[\vett{J}]$. In terms of the Feynman diagrams introduced above, in fact, $\mathcal{Z}_0[\vett{J}]$ can be written as follows:
%
%
\beq
Z_0[J] =
\begin{tikzpicture}
\begin{feynman}
\vertex [crossed dot] (a) {};
\vertex [above=0.7 cm  of a] (a0) {\(J_{\mu}(x)\)};
\vertex [crossed dot, right= of a] (b) {};
\vertex [above=0.7 cm  of b] (b0) {\(J_{\nu}(y)\)};
\diagram* {
(a0) (a) -- [boson] (b)
};
\end{feynman}
\end{tikzpicture}
+ \text{higher orders},
\eeq
where only the first order term in the $P$-expansion appearing in Eq. \eqref{feynman1} has been shown \cite{note4}, and each current node has been labelled with its correspondent current term, for later convenience.
%
%
Assume now, that we are interested in calculating the two-point correlation function, ie., the dressed photon propagator, $\langle A_{\mu}(x_1)A_{\nu}(x_2)\rangle$. According to the results presented in the previous section, this can be calculated via the relation
\beq
<A_{\mu}(\omega,\vett{x}_1) A_{\nu}(\omega,\vett{x}_2)>\propto\frac{\delta^2\mathcal{Z}_0[J]}{\delta J_{\mu}(\omega,\vett{x}_1)J_{\nu}(\omega,\vett{x}_2)}\Bigg|_{J=0}.
\eeq
In terms of Feynman diagrams, this can be understood as follows: every functional derivative in the equation above removes a source (crossed dot) from $\mathcal{Z}_0[\vett{J}]$, and labels the correspondent endpoint with the coordinate $x_{1,2}\equiv\{\omega,\vett{x}_{1,2}\}$, at which that specific functional derivative is taken, namely
%
%
\barr
&&<A_{\mu}(\omega,\vett{x}_1) A_{\nu}(\omega,\vett{x}_2)>=\frac{\delta^2\mathcal{Z}_0[\vett{J}]}{\delta J_{\mu}(\omega,\vett{x}_1)\delta J_{\nu}(\omega,\vett{x}_2)}\Bigg|_{J=0}\nonumber\\
&=&\frac{\delta^2}{\delta J_{\mu}(\omega,\vett{x}_1)\delta J_{\nu}(\omega,\vett{x}_2)}\Bigg\{
\begin{tikzpicture}
\begin{feynman}
\vertex [crossed dot] (a) {};
\vertex [above=0.7 cm  of a] (a0) {\(J_{\alpha}(x)\)};
\vertex [crossed dot, right= of a] (b) {};
\vertex [above=0.7 cm  of b] (b0) {\(J_{\beta}(y)\)};
\diagram* {
(a0) (a) -- [boson] (b)
};
\end{feynman}
\end{tikzpicture}
\Bigg\}\Bigg|_{J=0}\nonumber\\
&=&
\begin{tikzpicture}
\begin{feynman}
\vertex (a) {};
\vertex [above= 0.5 cm of a] (a0) { };
\vertex [right=of a] (b) {};
\vertex [ above= 0.5 cm of b] (b0) {};
\vertex [above left=0.4 cm of a] (at) {\((\omega, \vett{x}_1)\)};
\vertex[above right=0.4 cm of b] (bt) {\((\omega, \vett{x}_2)\)};

\diagram* {
(a0) (a) -- [boson] (b) (b0)
};
\end{feynman}
\end{tikzpicture}
\nonumber\\
&=& G_{\mu\nu}(\omega,\vett{x}_1-\vett{x}_2).
\earr
%
%
The end result of this calculation is, as expected, the dressed photon propagator $G_{\mu\nu}(\omega,\vett{x}_1-\vett{x}_2)$. This, then, describes the propagation of a single photon (of frequency $\omega$) inside the medium, from point $\vett{x}_1$ to point $\vett{x}_2$.
\subsection{Relevant diagrams for $\mathcal{Z}[\vett{J}]$}
We can now introduce the relevant diagrams for the partition function for $\chi^{(2)}$-processes. To do that, notice, that in writing Eq. \eqref{chi2Z}, we have assumed to consider only terms up to first order in $\lambda$. This corresponds to consider terms up to $V=1$, and $P=2$, in Eq. \eqref{feynman1}, i.e., Eq. \eqref{chi2Z} is described only by Feynman diagrams with $V=1$ vertices, and $P=2$ propagators (associated to the signal, and idler, mode, respectively). The fundamental Feynman diagrams for $\chi^{(2)}$-processes are shown in Fig. \ref{figure1}. As it can be seen, there are 3 relevant diagrams, describing the three basic $\chi^{(2)}$ processes of SPDC [Fig. \ref{figure1}(a)], i.e., the spontaneous generation of a signal-idler photon pair from the dressed vacuum, DFG [Fig. \ref{figure1}(b)], i.e., the scattering of a signal photon into an idler one, and SFG [Fig. \ref{figure1}(c)], namely the recombination of a signal-idler photon pair into vacuum (i.e., into a pump photon). However, as $\chi^{(2)}$ processes involve three photons, one would expect six different diagrams (as there are $3!=6$ different ways to arrange the three different diagrams appearing in Fig. \ref{figure1}). The missing three diagrams, can be easily obtained from the ones depicted in Fig. \ref{figure1}, by exchanging the role of the signal and idler modes. Moreover, notice, that SHG is a special case of Fig. \ref{figure1}(c), when the signal and idler photons are degenerate, i.e., $\omega_s=\omega_i=\omega_p/2$.

Notice, moreover, that the diagrams in Fig. \ref{figure1} also possess an extra symmetry, deriving from the possibility, to exchange the current endpoints in each diagram, without changing the meaning of the diagram itself. This amounts to an extra three equivalent diagrams.

In terms of Feynman diagrams, then, the partition function for $\chi^{(2)}$ processes can be written, at the order $\mathcal{O}(\lambda)$, as follows:
\begin{widetext}
\beq
\mathcal{Z}[\vett{J}]=
\begin{gathered}
\begin{tikzpicture}
\begin{feynman}
\vertex  (a0){};
\vertex [dot, right= of a0] (a){};
\vertex [crossed dot, above right= of a] (w1){};
\vertex [right = 0.7 cm of w1] (c1) {};
\vertex [crossed dot, below right= of a] (w2){};
\vertex [right = 0.7 cm of w2] (c2) {};
\diagram* {
(a0) -- [scalar] (a) -- [boson, edge label= \(\omega_s\)] (w1),
(a) -- [boson, edge label' = \(\omega_i\)] (w2),
};
\end{feynman}
\end{tikzpicture}
\end{gathered}
+\hspace{1mm}
\begin{gathered}
\begin{tikzpicture}
\begin{feynman}
\vertex [crossed dot] (a0){};
\vertex [above = 0.4 cm of a0] (c1) {};
\vertex [dot, right= of a0] (a){};
\vertex [crossed dot, above right= of a] (w1){};
\vertex [below right= of a] (w2){};
\vertex [right = 0.7 cm of w1] (c2) {};
\diagram* {
(a0) --  [boson, edge label= \(\omega_s\)]  (a) --[boson, edge label= \(\omega_i\)]  (w1),
(a) --  [scalar]  (w2),
};
\end{feynman}
\end{tikzpicture}
\end{gathered}
+\hspace{1mm}
\begin{gathered}
\begin{tikzpicture}
\begin{feynman}
\vertex [dot] (a0){};
\vertex [crossed dot, above left= of a0] (b) {};
\vertex [crossed dot, below left= of a0] (c) {};
\vertex [right= of a0] (d) {};
\vertex [above=0.4cm of b](c1) {};
\vertex [below=0.4cm of c](c2) {};
\diagram* {
(a0) -- [boson, edge label'= \(\omega_i\)] (c),
(a0) -- [boson, edge label'= \(\omega_s\)] (b),
(a0) -- [scalar] (d),
};
\end{feynman}
\end{tikzpicture}
\end{gathered}
+ \text{$\omega_s\leftrightarrow\omega_i$}
\eeq
\end{widetext}
\subsection{Cross section for $\chi^{(2)}$ processes}
From a physical point of view, the processes depicted in Fig. \ref{figure1} are very different. SPDC, for example, is a spontaneous process, originating from the dressed vacuum. DFG and SFG, on the other hand, require the pre-existing presence of signal and/or idler photons, that can seed the process, and make it possible. These seeds can be either present in the initial state of the electromagnetic field (i.e., a weak signal (idler) pulse, which impinges into the nonlinear medium, together with the pump beam), or they can appear as a result of cascaded nonlinear processes in the medium itself, which are typically described by higher order interactions [i.e., $V>1$ terms in Eq. \eqref{feynman1}]. The initial state of the electromagnetic field before the nonlinear interaction (or, in case of cascaded processes, the types of allowed higher order interactions) plays then an important role, to select out, which processes can take place inside the medium.

Despite the physical and conceptual difference between the various nonlinear processes described by $\chi^{(2)}$-nonlinearities, however, their cross section is the same for any of such processes, and proportional to the two-point correlation function $\langle A_{\mu}(x)A_{\nu}(y)\rangle$. The reason for this resides in the fact, that, in the undepleted pump approximation, the interaction Lagrangian describing $\chi^{(2)}$-processes is quadratic in the vector potential. Within the undepleted pump approximation, therefore, nonlinear interactions can be described in terms of effective , self-interacting free fields, for which, as it is well known in quantum field theory \cite{das}, the only relevant quantity is the two-point correlation function.

For the case of SPDC, for example, we have
\barr\label{eq203}
\sigma_{SPDC}&=&<A_{\mu}(x)A_{\nu}(y)>=\frac{\delta^2\mathcal{Z}[\vett{J}]}{\delta J_{\mu}(x)\delta J_{\nu}(y)}\Bigg|_{\vett{J}=0}\nonumber\\
&=&
\begin{gathered}
\begin{tikzpicture}
\begin{feynman}
\vertex  (a0){};
\vertex [dot, right= of a0] (a){};
\vertex [above right= of a] (w1){};
\vertex [above= 0.5 cm of w1] (w10) {};
\vertex [below right= of a] (w2){};
\vertex [below= 0.5 cm of w2] (w20) {};
\vertex [above right=0.1 cm of w1] (w1t) {\(x\)};
\vertex [below right=0.1 cm of w2] (w2t) {\(y\)};
\diagram* {
(a0) -- [scalar] (a) -- [boson, edge label=\(\omega_s\)] (w1),
(a) -- [boson, edge label'=\(\omega_i\)] (w2),
};
\end{feynman}
\end{tikzpicture}
\end{gathered}
=\frac{1}{\hbar^2}\mathcal{X}_{\mu\nu}^{(2)}(x-y)\nonumber\\
&=&\frac{1}{\hbar^2}\int d^4z\lambda_{\alpha\beta}(z)G_{\mu\alpha}(x-z)G_{\beta\nu}(z-y),
\earr
where $x=\{\omega_s,\vett{x}\}$, and $y=\{\omega_i,\vett{y}\}$

For DFG, on the other hand, we have
\barr\label{eq203}
\sigma_{DFG}&=&<A_{\mu}(x)A_{\nu}(y)>=\frac{\delta^2\mathcal{Z}[\vett{J}]}{\delta J_{\mu}(x)\delta J_{\nu}(y)}\Bigg|_{\vett{J}=0}\nonumber\\
&=&
\begin{gathered}
\begin{tikzpicture}
\begin{feynman}
\vertex (a0){};
\vertex [above = 0.4 cm of a0] (c1) {};
\vertex [dot, right= of a0] (a){};
\vertex [above right= of a] (w1){};
\vertex [below right= of a] (w2){};
\vertex [right = 0.7 cm of w1] (c2) {};
\vertex [above left=0.1 cm of a0] (w1t) {\(x\)};
\vertex [below right=0.1 cm of w1] (w2t) {\(y\)};
\diagram* {
(a0) --  [boson, edge label= \(\omega_s\)]  (a) --[boson, edge label= \(\omega_i\)]  (w1),
(a) --  [scalar]  (w2),
};
\end{feynman}
\end{tikzpicture}
\end{gathered}
=\frac{1}{\hbar^2}\mathcal{X}_{\mu\nu}^{(2)}(x-y)\nonumber\\
&=&\frac{1}{\hbar^2}\int d^4z\lambda_{\alpha\beta}(z)G_{\mu\alpha}(x-z)G_{\beta\nu}(z-y).
\earr
An analogous result can be obtained for the cross section of SFG. 

This, in principle, states, that, ideally, SPDC, SFG, and SFG, despite being three different physical nonlinear processes, have the same likelihood to happen, inside a nonlinear medium. This is not surprising, since the diagrams in Fig. \ref{figure1} can be easily transformed into one another, and are therefore equivalent. Again, the initial state of the electromagnetic field will define, which processes will take place. If we assume, for example, that the electromagnetic field is initially in its dressed vacuum state (i.e., only the pump beam impinges onto the nonlinear medium), then, the only possible process would be SPDC, as it is the only Feynman diagram, capable of generating a photon pair out of vacuum  [Fig. 1(a)]. If, instead, we assume, that the electromagnetic field is, say, initially in the state $\ket{1_s}$, then, not only SPDC (which, it being a spontaneous process, is always present) will be possible, but also the scattering processes, that are described by Feynman diagrams, which convert a signal photon, into an idler one [Fig. 1(b)]. The same is valid, if the initial state of the electromagnetic field would be $\ket{1_i}$, instead. 

It is worth noticing, that this line of reasoning is valid, when the electromagnetic field is in an initial state containing exactly one signal (idler) photon. If we consider a more realistic situation, where the signal (idler) field is described by a weak coherent state $\ket{\alpha_s}$, with $|\alpha_s|^2\ll 1$, then we are in the situation, that the cross section for DFG events will be $|\alpha_s|^2$ times larger, than the SPDC one \cite{quantumNLO}. This, again, is not surprising, since, within our theory, DFG is a stimulated process, while SPDC is a spontaneous one.

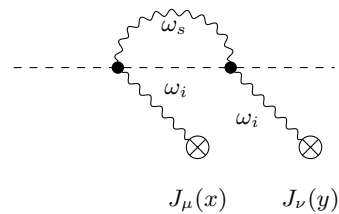
\begin{figure}[t!]
\begin{tikzpicture}
\begin{feynman}
\vertex (a0) {};
\vertex  [dot,right=of a0](a) {};
\vertex [dot, right=of a](c){};
\vertex [crossed dot, below right=of a] (w1){};
\vertex [below=0.7 cm of w1] (w10) {\(J_{\mu}(x)\)};
\vertex [crossed dot, below right=of c] (w2){};
\vertex [below=0.7 cm of w2] (w20) {\(J_{\nu}(y)\)};
\vertex [right=of c] (d){};
\diagram* {
(a0) -- [scalar] (a) -- [boson, half left, edge label'=\(\omega_s\)] (c) -- [boson, edge label'=\(\omega_i\)] (w2),
(a) -- [scalar] (c),
(a) --  [boson, edge label=\(\omega_i\)] (w1),
(c) -- [scalar] (d),
};
\end{feynman}
\end{tikzpicture}
\caption{Higher order Feynman diagram depicting the process $\ket{0}\rightarrow\ket{1_s,1_i}\rightarrow\ket{2_i}$. First, a signal-idler photon pair is created from the dressed vacuum (first interaction point), as described by Fig. \ref{figure1}(a). Then, since the state of the system after this interaction is given by $\ket{1_s,1_i}$, a second first-order process (second interaction point), namely the scattering of a signal photon into an idler one [Fig. \ref{figure1}(b)] might take place. However, since this process involves $V=2$ vertices, it is a process of order $\mathcal{O}(\lambda^2)$, and therefore not present in the expression of $\mathcal{Z}[\vett{J}]$, which only contains processes up to $\mathcal{O}(\lambda)$.}
\label{figure2}
\end{figure}

As a last remark, notice, that since $\mathcal{Z}[\vett{J}]$ only contains first-order terms in the $V$-expansion of Eq. \eqref{feynman1}, these processes depicted in Fig. \ref{figure1} are the only possible processes. However, in a real experiment, cascaded processes, such as, for example, $\ket{0}\rightarrow\ket{1_s,1_i}\rightarrow\ket{2_i}$, may be triggered. These processes, however, are described by a two-vertex Feynman diagram, as the one depicted in Fig. \ref{figure2}, and correspond to terms of order $\mathcal{O}(\lambda^2)$, which are neglected in Eq. \eqref{chi2Z}.
%
%
%
%

%
%
\subsection{Cascaded $\chi^{(2)}$ processes}
If we want to account for higher order processes, like the one depicted in Fig. \ref{figure2}, we need to expand the expression of $\mathcal{Z}[\vett{J}]$, so that it accounts also for higher powers of the coupling constant $\lambda$. These processes, in optics, are known as cascaded processes. To describe such processes within the framework developed in the previous section, we need to write the partition function as $\mathcal{Z}[\vett{J}]=\mathcal{Z}_0[\vett{J}]+\mathcal{Z}_1[\vett{J}]+\mathcal{Z}_2[\vett{J}]+\mathcal{O}(\lambda^3)$, where $\mathcal{Z}_2[\vett{J}]$ accounts for the occurrence of nonlinear interactions with 2 vertices, whose explicit expresison can be derived analytically from Eq. \eqref{intFinal}, by keeping terms up to $n=2$, and then use this result into Eq. \eqref{feynman1}, by keeping terms up to $V=2$. The explicit expression of $\mathcal{Z}_2[\vett{J}]$, however, is rather cumbersome, but it can be written easily written in terms of Feynman diagrams as follows:
\begin{widetext}
\barr\label{Z2diag}
\mathcal{Z}_2[\vett{J}]&=&
\begin{gathered}
\begin{tikzpicture}
\begin{feynman}
\vertex (a0) {};
\vertex  [dot,right=of a0](a) {};
\vertex [dot, right=of a](c){};
\vertex [above=0.7 cm of w1] (w10) {};
\vertex [below=0.7 cm of w2] (w20) {};
\vertex [right=of c] (d){};
\diagram* {
(a0) -- [scalar] (a) -- [boson, half left, edge label'=\(\omega_s\)] (c)
(c) -- [boson, half left, edge label'=\(\omega_i\)] (a)
(a) -- [scalar] (c),
(c) -- [scalar] (d),
};
\end{feynman}
\end{tikzpicture}
\end{gathered}
+\hspace{1mm}
\begin{gathered}
\begin{tikzpicture}
\begin{feynman}
\vertex (a0) {};
\vertex  [dot,right=of a0](a) {};
\vertex [dot, right=of a](c){};
\vertex [crossed dot, below right=of a] (w1){};
\vertex [above=0.7 cm of w1] (w10) {};
\vertex [crossed dot, below right=of c] (w2){};
\vertex [below=0.7 cm of w2] (w20) {};
\vertex [right=of c] (d){};
\diagram* {
(a0) -- [scalar] (a) -- [boson, half left, edge label'=\(\omega_s\)] (c) -- [boson, edge label'=\(\omega_i\)] (w2),
(a) -- [scalar] (c),
(a) --  [boson, edge label=\(\omega_i\)] (w1),
(c) -- [scalar] (d),
};
\end{feynman}
\end{tikzpicture}
\end{gathered}
\nonumber\\
&+&
\begin{gathered}
\begin{tikzpicture}
\begin{feynman}
\vertex  (a0){};
\vertex [dot, right= of a0] (a){};
\vertex [crossed dot, above right= of a] (w1){};
\vertex [right = 0.7 cm of w1] (c1) {};
\vertex [crossed dot, below right= of a] (w2){};
\vertex [right = 0.7 cm of w2] (c2) {};
\vertex [dot, right= of a] (d) {};
\vertex [crossed dot, above right= of d] (w11){};
\vertex [crossed dot, below right= of d] (w21){};
\vertex [right= of d] (e) {};
\diagram* {
(a0) -- [scalar] (a) -- [boson, edge label= \(\omega_s\)] (w1),
(a) -- [boson, edge label' = \(\omega_i\)] (w2),
(a) -- [scalar] (d) -- [boson, edge label= \(\omega_s\)] (w11),
(d) -- [boson, edge label' = \(\omega_i\)] (w21),
(d) -- [scalar] (e),
};
\end{feynman}
\end{tikzpicture}
\end{gathered}
+\hspace{1mm}
\begin{gathered}
\begin{tikzpicture}
\begin{feynman}
\vertex [dot] (a0){};
\vertex [crossed dot, above left= of a0] (b) {};
\vertex [crossed dot, below left= of a0] (c) {};
\vertex [dot, right= of a0] (d) {};
\vertex [crossed dot, above right= of d] (b1) {};
\vertex [crossed dot, below right= of d] (c1) {};
\vertex [right= of d] (e) {};

\diagram* {
(a0) -- [boson, edge label'= \(\omega_i\)] (c),
(a0) -- [boson, edge label'= \(\omega_s\)] (b),
(a0) -- [scalar] (d),
(d) -- [boson, edge label'= \(\omega_i\)] (c1),
(d) -- [boson, edge label'= \(\omega_s\)] (b1),
(d) -- [scalar] (e),
};
\end{feynman}
\end{tikzpicture}
\end{gathered}
\nonumber\\
&+&
\begin{gathered}
\begin{tikzpicture}
\begin{feynman}
\vertex [crossed dot] (a0){};
\vertex [dot, right= of a0] (a){};
\vertex [dot, above right= of a] (w1){};
\vertex [below right= of a] (w2){};
\vertex [right= of w1] (d) {};
\vertex [crossed dot, above left= of w1] (e) {};
\diagram* {
(a0) --  [boson, edge label= \(\omega_s\)]  (a) --[boson, edge label= \(\omega_i\)]  (w1),
(a) --  [scalar]  (w2),
(w1) -- [scalar] (d),
(e) --  [boson, edge label= \(\omega_s\)]  (w1),
};
\end{feynman}
\end{tikzpicture}
\end{gathered}
+\hspace{1mm}
\begin{gathered}
\begin{tikzpicture}
\begin{feynman}
\vertex [crossed dot] (a0){};
\vertex [dot, right= of a0] (a){};
\vertex [dot, above right= of a] (w1){};
\vertex [below right= of a] (w2){};
\vertex [crossed dot, right= of w1] (d) {};
\vertex [above left= of w1] (e) {};
\diagram* {
(a0) --  [boson, edge label= \(\omega_s\)]  (a) --[boson, edge label= \(\omega_i\)]  (w1),
(a) --  [scalar]  (w2),
(w1) --  [boson, edge label= \(\omega_s\)]  (d),
(e) --  [scalar] (w1),
};
\end{feynman}
\end{tikzpicture}
\end{gathered}
+\text{$\omega_s\leftrightarrow\omega_i$}
\earr
\end{widetext}
Notice, that although the first diagram in the above equation contains a loop, this loop gives a finite result, as it corresponds to the process of SPDC, followed by SFG, which, essentially, cancel each other, leaving only photons in the pump mode.

Contrary to first order processes, which, as we discussed above, have all the same cross section, in this case, the cross section is different for different processes. A careful analysis of the diagrams presented above, in fact, reveals, that they can be grouped into two groups, namely those diagrams containing  two current sources, and those containing four. Processes belonging to these two classes will have different cross sections. Ultimately, this is due to the extra symmetry given by the presence of the current sources, and the fact, that by interchanging these sources, the resulting diagram will be equivalent to the original one. However, diagrams with only 2 sources will have an extra factor of 2 in their cross section (as there are $2!=2$ equivalent diagrams, corresponding to the two different way to arrange two current sources), while the processes containing four current sources will have an extra factor of 24, as we can arrange the four current sources in $4!=24$ different ways.

 As an example, we report the cross section for the cascaded SPDC process, i.e., for the third diagram appearing in Eq. \eqref{Z2diag}
\barr
\sigma^{casc}_{SPDC}&=& 
\begin{gathered}
\begin{tikzpicture}
\begin{feynman}
\vertex  (a0){};
\vertex [dot, right= of a0] (a){};
\vertex [above right= of a] (w1){};
\vertex [right = 0.7 cm of w1] (c1) {};
\vertex [below right= of a] (w2){};
\vertex [right = 0.7 cm of w2] (c2) {};
\vertex [dot, right= of a] (d) {};
\vertex [above right= of d] (w11){};
\vertex [below right= of d] (w21){};
\vertex [right= of d] (e) {};
\vertex [above right= 0.1 cm of w1] (w1t) {\(x_1\)};
\vertex [above right=0.1 cm of w11] (w11t) {\(x_3\)};
\vertex [below right=0.11 cm of w2] (w2t) {\(x_2\)};
\vertex [below right=0.1 cm of w21] (w21t) {\(x_4\)};
\diagram* {
(a0) -- [scalar] (a) -- [boson, edge label= \(\omega_s\)] (w1),
(a) -- [boson, edge label' = \(\omega_i\)] (w2),
(a) -- [scalar] (d) -- [boson, edge label= \(\omega_s\)] (w11),
(d) -- [boson, edge label' = \(\omega_i\)] (w21),
(d) -- [scalar] (e),
};
\end{feynman}
\end{tikzpicture}
\end{gathered}
\nonumber\\
&=&\frac{24}{\hbar^4}\mathcal{X}_{\mu\nu}^{(2)}(x_1-x_2)\mathcal{X}_{\alpha\beta}^{(2)}(x_3-x_4),
\earr
where $x_{1,2}=\{\omega_{s,i},\vett{x}_{1,2}\}$ are the coordinates associated to the first signal-idler photon pair, while $x_{3,4}=\{\omega_{s,i},\vett{x}_{3,4}\}$ are associated to the second signal-idler photon pair. In general, $x_{1,2}\neq x_{3,4}$, since the two photon pair might be generated at a slightly different frequency (especially, if a broadband is used for this process), or they might be generated in different points inside the nonlinear medium.
In the special case, in which the signal and idler modes are degenerate, the above expression simplifies to $\sigma^{casc}_{SPDC}=24\left[\sigma_{SPDC}\right]^2$. This result will be useful, when describing the generation of squeezed light.
\section{Some Examples}
In this section, we apply the formalism developed above to two simple examples, namely the occurrence of SPDC in a one dimensional, homogeneous, nonlinear, lossy medium, and the generation of squeezed light from repeated cascaded SPDC processes.
\subsection{SPDC in a one dimensional lossy waveguide}
Let us consider a one dimensional, homogeneous, nonlinear, medium. For the sake of simplicity, we could imagine this medium to be an optical waveguide of length $L$ along the $x$-direction, characterised by a refractive index $n(\omega)$, and a nonlinear susceptibility $\chi^{(2)}_{\mu\nu\sigma}\equiv\chi$, since we have assumed the medium to be homogeneous. With this definition, we can introduce the signal, and idler wave vectors in the medium, as $k_{s,i}(\omega)=k_0n_r(\omega_{s,i})\equiv k_{s,i}$. Moreover, if we assume, that the pump beam can be written as a plane wave, the nonlinear coupling constant appearing in Eq. \eqref{eq203} assumes a simpler expression, namely $\lambda_{\alpha\beta}(z)=\chi A_p\exp{[i(k_px-\omega_pt)]}$, and, in particular, is independent on $z$.

According to the results presented in Sect. V.B, the cross section associated to the generation of a signal-idler photon pair from an SPDC event, assuming the initial state of the effective electromagnetic field to be $\ket{0}$ is given by Eq. \eqref{eq203}, i.e.,
\beq
\sigma_{SPDC}=\frac{1}{\hbar^2}\mathcal{X}^{(2)}_{\mu\nu}(x-y),
\eeq
where the biphoton propagator is given by
\beq\label{eq48}
\mathcal{X}^{(2)}_{\mu\nu}(x-y)=\int d^4zG_{\mu\alpha}(x-z)\lambda_{\alpha\beta}(z)G_{\beta\nu}(z-y).
\eeq
For a one dimensional, homogeneous medium, the dressed photon propagator can be easily calculated from Eq. \eqref{helmholtz}, which, in this case, reduces to
\beq
\left[\frac{\partial^2}{\partial x^2}+k_0^2n^2(\omega)\right]G(\omega,x-y)=\delta(x-y),
\eeq
where $k_0=\omega/c$ is the vacuum wave vector, and the explicit expression of the dressed propagator is given by \cite{das}
\barr
G(\omega,x-y)&=&\frac{1}{2ik(\omega)}\Big[\Theta(x-y)e^{i[k(\omega)(x-y)-\omega t]}\nonumber\\
&+&\Theta(y-x)e^{-i[k(\omega)(x-y)-\omega t]}\Big],
\earr
where $\Theta(x)$ is the Heaviside step function \cite{nist}. Moreover, of the two propagators appearing in Eq. \eqref{eq48}, one describes the dynamics of the signal field, and must be then evaluated at the signal frequency $\omega_s$, while the other describes the dynamics of the idler field, and must then be evaluated at the idler frequency $\omega_i$.

We can now calculate the explicit expression of the biphoton propagator $\mathcal{X}^{(2)}_{\mu\nu}(x-y)$, which in this casse is given by
\barr\label{biphoton1D}
\mathcal{X}^{(2)}(x-y)&=&\Theta(x-y)\mathcal{G}(x,y)e^{i\Delta\omega t}L\sinc\left(\frac{L\Delta k}{2}\right)\nonumber\\
&+&\text{phase mismatched terms},
\earr
where $\Delta\omega=\omega_p-\omega_s-\omega_i$ is the frequency mismatch, constrained to be zero by energy conservation, i.e., by $\omega_p=\omega_s+\omega_i$ \cite{boyd}, $\Delta k=k_p+k_s+k_i$ is the phase mismatch, and
\beq
\mathcal{G}(x,y)=\frac{\chi A_p}{4k_sk_i}e^{-i(k_sx+k_iy)}.
\eeq
The label ``\emph{phase mismatched terms}" in the above equation, moreover, refers to those terms in the expression of $\mathcal{X}^{(2)}(x-y)$, that violate either the energy, or the momentum conservation laws, and that are therefore forbidden.

Notice, moreover, that since nonlinear processes conserve not only energy, but also momentum, $\Delta k=0$ must be also fulfilled. This means, that $k_p=-(k_s+k_i)$, i.e., that the only possibility for SPDC to take place in a one dimensional system is, that the signal and the idler photons are emitted in the opposite direction, with respect to the direction of propagation of the pump beam.

The probability for a SPDC event to occur is then given by
\beq
P(L)\propto|\sigma_{SPDC}|^2\propto L^2\sinc^2\left(\frac{L\Delta k}{2}\right),
\eeq
which is in accordance with standard results \cite{boyd}.
\subsection{Squeezing}
In this second example, we consider the situation of the occurrence, inside a nonlinear medium, of $N$ cascaded SPDC processes, and we investigate, how this process can be connected to squeezing. To make things easier, let us assume, that SPDC is degenerate, i.e., that the signal-idler photon pair has the same frequency, namely $\omega_s=\omega_i\equiv\Omega$, and it gets created in the same frequency mode. Since we are only interested in SPDC events, we can consider only the part of the partition function, that accounts for SPDC solely. In a real experiment, of course, this would not be the case, as many other nonlinear effect will take place concurrently with SPDC, but for the sake of this discussion, we can neglect this fact. The partition function describing this process then contains all the SPDC events, up to order $N$ in the expansion in power series of $\lambda$, i.e,
\begin{widetext}
\barr
\mathcal{Z}[\vett{J}]&=&
\begin{gathered}
\begin{tikzpicture}
\begin{feynman}
\vertex  (a0){};
\vertex [dot, right= of a0] (a){};
\vertex [crossed dot, above right= of a] (w1){};
\vertex [right = 0.7 cm of w1] (c1) {};
\vertex [crossed dot, below right= of a] (w2){};
\vertex [right = 0.7 cm of w2] (c2) {};
\diagram* {
(a0) -- [scalar] (a) -- [boson, edge label= \(\Omega\)] (w1),
(a) -- [boson, edge label' = \(\Omega\)] (w2),
};
\end{feynman}
\end{tikzpicture}
\end{gathered}
+\hspace{1mm}
\begin{gathered}
\begin{tikzpicture}
\begin{feynman}
\vertex  (a0){};
\vertex [dot, right= of a0] (a){};
\vertex [crossed dot, above right= of a] (w1){};
\vertex [right = 0.7 cm of w1] (c1) {};
\vertex [crossed dot, below right= of a] (w2){};
\vertex [right = 0.7 cm of w2] (c2) {};
\vertex [dot, right= of a] (d) {};
\vertex [crossed dot, above right= of d] (w11){};
\vertex [crossed dot, below right= of d] (w21){};
\vertex [right= of d] (e) {};
\diagram* {
(a0) -- [scalar] (a) -- [boson, edge label= \(\Omega\)] (w1),
(a) -- [boson, edge label' = \(\Omega\)] (w2),
(a) -- [scalar] (d) -- [boson, edge label= \(\Omega\)] (w11),
(d) -- [boson, edge label' = \(\Omega\)] (w21),
(d) -- [scalar] (e),
};
\end{feynman}
\end{tikzpicture}
\end{gathered}
\nonumber\\ &+&
\begin{gathered}
\begin{tikzpicture}
\begin{feynman}
\vertex  (a0){};
\vertex [dot, right= of a0] (a){};
\vertex [crossed dot, above right= of a] (w1){};
\vertex [right = 0.7 cm of w1] (c1) {};
\vertex [crossed dot, below right= of a] (w2){};
\vertex [right = 0.7 cm of w2] (c2) {};
\vertex [dot, right= of a] (d) {};
\vertex [crossed dot, above right= of d] (w11){};
\vertex [crossed dot, below right= of d] (w21){};
\vertex [dot, right= of d] (e) {};
\vertex [crossed dot, above right= of e] (w12){};
\vertex [crossed dot, below right= of e] (w22){};
\vertex [right= of e] (f) {};
\diagram* {
(a0) -- [scalar] (a) -- [boson, edge label= \(\Omega\)] (w1),
(a) -- [boson, edge label' = \(\Omega\)] (w2),
(a) -- [scalar] (d) -- [boson, edge label= \(\Omega\)] (w11),
(d) -- [boson, edge label' = \(\Omega\)] (w21),
(d) -- [scalar] (e) --[boson, edge label= \(\Omega\)] (w12),
(e) -- [boson, edge label' = \(\Omega\)] (w22),
(e) -- [scalar] (f),
};
\end{feynman}
\end{tikzpicture}
\end{gathered}
\nonumber\\ &+&
\begin{gathered}
\begin{tikzpicture}
\begin{feynman}
\vertex  (a0){};
\vertex [dot, right= of a0] (a){};
\vertex [crossed dot, above right= of a] (w1){};
\vertex [right = 0.7 cm of w1] (c1) {};
\vertex [crossed dot, below right= of a] (w2){};
\vertex [right = 0.7 cm of w2] (c2) {};
\vertex [dot, right= of a] (d) {};
\vertex [crossed dot, above right= of d] (w11){};
\vertex [crossed dot, below right= of d] (w21){};
\vertex [right= of d] (e) {};
\diagram* {
(a0) -- [scalar] (a) -- [boson, edge label= \(\Omega\)] (w1),
(a) -- [boson, edge label' = \(\Omega\)] (w2),
(a) -- [scalar] (d) -- [boson, edge label= \(\Omega\)] (w11),
(d) -- [boson, edge label' = \(\Omega\)] (w21),
(d) -- [scalar] (e),
};
\end{feynman}
\end{tikzpicture}
\end{gathered}
\cdots\hspace{1mm}
\begin{gathered}
\begin{tikzpicture}
\begin{feynman}
\vertex  (a0){};
\vertex [dot, right= of a0] (a){};
\vertex [crossed dot, above right= of a] (w1){};
\vertex [right = 0.7 cm of w1] (c1) {};
\vertex [crossed dot, below right= of a] (w2){};
\vertex [right = 0.7 cm of w2] (c2) {};
\vertex [right= of a] (b);
\diagram* {
(a0) -- [scalar] (a) -- [boson, edge label= \(\Omega\)] (w1),
(a) -- [boson, edge label' = \(\Omega\)] (w2),
(a) -- [scalar] (b),
};
\end{feynman}
\end{tikzpicture}
\end{gathered}
,
\earr
\end{widetext}
where the last diagram contains $N$ cascaded SPDC processes. The diagrams shown above, describe processes of the type $\ket{0}\rightarrow\ket{0}+\ket{2}\rightarrow\ket{0}+\ket{2}+\ket{4}\rightarrow\cdots\rightarrow\ket{0}+\ket{2}+\ket{4}+\cdots+\ket{2N}$.

As we. have seen in the previous section, for cascaded, degenerated SPDC, we have, that $\sigma_{SPDC}^{casc}\propto\sigma_{SPDC}^2$. Hence, we can generalise this result, to an arbitrary N-fold cascaded process, by simply saying, that $\sigma_{SPDC}^{(N)}=\sigma_{SPDC}^{N/2}$. If we now let $N\rightarrow\infty$, and reconstruct the final state of the electromagnetic field, as the sum of all these interactions, we obtain
\beq\label{spdcS}
\ket{\psi}=\sum_{k\in\{even\}}\psi_{k}\left(\sigma_{SPDC}\right)^k\ket{k},
\eeq
where $\psi_{k}$ is ja suitable normalisation constant, chosen in such a way, that $\braket{\psi}{\psi}=1$.  A closer inspection on $\ket{\psi}$ reveals, that it only contains modes with an even number of photons in them. This is the typical form of a single mode squeezed state \cite{loudon}, i.e.,
\beq\label{squeezing}
\ket{\xi}=\sqrt{\sech s}\sum_{k\in\{even\}}\sqrt{\frac{(2k)!}{k!}}\left(-\frac{1}{2}e^{i\theta}\tanh s\right)^k\ket{k},
\eeq
where $\xi=s\exp{(i\theta)}$ is the squeezing parameter. 

If we compare Eq. \eqref{spdcS} with Eq. \eqref{squeezing}, we can relate, up to a normalisation constant, the squeezing parameter $\xi$ with the SPDC cross section, and, therefore, with the properties of the nonlinear medium. We then have
\beq\label{eq56}
\psi_k\sigma_{SPDC}^k=\sqrt{\sech s\frac{(2k)!}{k!}}\left(-\frac{1}{2}\right)^k\tanh^k s e^{ik\theta}.
\eeq
If we now call $\sigma_{SPDC}=\rho_{\sigma}e^{i\varphi_{\sigma}}$, where, according to Eq. \eqref{eq203},
\bseq
\begin{align}
\rho_{\sigma}&=\frac{1}{\hbar^2}\left|\mathcal{X}_{\mu\nu}^{(2)}(x-y)\right|,\\
\varphi_{\sigma}&=\operatorname{Arg}\left\{\mathcal{X}_{\mu\nu}^{(2)}(x-y)\right\},
\end{align}
\eseq
and substitute these expression into Eq. \eqref{eq56} we have, up to a normalisation constant,
\bseq\label{squeezingMat}
\begin{align}
\tanh s &= \rho_{sigma}=\frac{1}{\hbar^2}\left|\mathcal{X}_{\mu\nu}^{(2)}(x-y)\right|,\\
\theta &=\varphi_{\sigma}= \operatorname{Arg}\left\{\mathcal{X}_{\mu\nu}^{(2)}(x-y)\right\}.
 \end{align}
\eseq
For the particular case of one dimensional SPDC treated above, we can analytically calculate the modulus and phase of the squeezing parameter using Eq. \eqref{biphoton1D}, which gives, in case of perfect phase matching (i.e., $\Delta k=0$), and assuming, that $A_p=|A_p|\exp{(i\phi_p)}$, the following result:
\bseq
\begin{align}
s&=\ln\sqrt{\frac{4k_sk_i+\chi |A_p| L}{4k_sk_i-\chi |A_p|L}},\\
\theta&=\phi_p-(k_sx+k_iy).
\end{align}
\eseq
This is an interesting result, because it generalises the standard results concerning single mode squeezing, where the properties of the pump beam define the squeezing parameter $\xi$ \cite{loudon}. In particular, we see from the above equations, that while the squeezing strength $s$ only depends on the pump amplitude, and the nonlinear coefficient $\chi$, the squeezing phase depends on the phase of the pump beam and, surprisingly, on the position at which the signal and idler photons are actually detected. 
\section{Summary}
In this work, we have used the formalism of path integrals, to describe the properties of the (quantised) electromagnetic field in an arbitrary nonlinear medium. Starting from the well known microscopic model investigated by Huttner and Barnett \cite{Huttner}, we have obtained the effective action \eqref{effectiveAction}, i.e.,
\barr
&&S_{eff}[\vett{A}]=S_{em}[\vett{A}]\nonumber\\
&+&\frac{1}{2}\int\,dt\,dt'\,d^3x\,g(\vett{x})\dot{\vett{A}}(t,\vett{x})\Gamma(t-t',\vett{x})\dot{\vett{A}}(t',\vett{x}),
\earr
and we have shown, that $\Gamma(t-t',\vett{x})$ contains all the informations about the medium, and can be interpreted as the Fourier transform of the dielectric constant of the medium itself. This result has allowed us, to describe the linear dynamics of the electromagnetic field in an arbitrary medium, as the ones of an effective free field, dressed by the presence of the medium. We then used the effective action above, to calculate the classical [Eq. \eqref{eq6}] and quantum [Eq. \eqref{freeTheory1}] partition function for the effective free theory. In particular, we have shown, that for the quantum free theory, the main quantity of interest is the Fourier transform of the dressed photon propagator, i.e., $G_{\mu\nu}(\omega,\vett{x}-\vett{x}')$.

In the second part of our manuscript, we concentrated on developing an interacting theory, describing the nonlinear interaction of the electromagnetic field in an arbitrary medium. By using standard results of quantum field theory, and the formalism of path integrals, we have shown, for second-order nonlinear processes in the undepleted pump approximation, that the quantity, that regulates the cross section of the various nonlinear processes is the biphoton propagator $\mathcal{X}_{\mu\nu}^{(2)}(x-y)$,i.e, 
\beq
\mathcal{X}^{(2)}_{\mu\nu}(x-y)=\int\,d^4z G_{\mu\alpha}(x-z)\lambda_{\alpha\beta}(z)G_{\beta\nu}(z-y).
\eeq
This is the main result of our work: the cross section (and, hence, the probability) of a second-ordeer process to happen inside an arbitrary nonlinear medium, is given, essentially, by the biphotn propagator defined above, which contains informations about both the signal, and the idler photons generated by the nonlinear processes, and individually described by their respective dressed propagators. This result is quite interesting, as it generalises the usual concept of biphoton wavefunction, frequently used in nonlinear optics \cite{quantumNLO,loudon}. 

Moreover, we have pointed out, how second-order nonlinear phenomena in the undepleted pump approximation can be treated, de facto, as effective free theories, as their correspndent integraction Lagrangian is quadratic in the electromagnetic field.

Last, we have presented two examples of application of our framework. In the first example, we have derived the usual $\sinc^2$-law for the generation of SPDC in a one dimensional crystal, and we have illustrated, how to calculate explicitly the biphoton propagator in such a simple example. 

Then, we have applied our result to a more complicated problem, namely the onset of progressive cascaded SPDC, and the onset of squeezing, and we have shown, how the familiar squeezing parameter $\xi=s\exp{(i\theta})$ can be linked to the linear, and, especially, nonlinear, properties of the medium [see Eqs. \eqref{squeezingMat}]. For the particular case of SPDC in a one dimensional system, moreover, we have presented an explicit expression of the form of the modulus and phase of the squeezing parameter, as a function of the properties of the medium, and the dynamics of the signal and idler field.

\section{Conclusions and Outlook}
In conclusion, our work presents a complete toolkit, based on the method of path integrals and Feynman diagrams, for calculating the classical and quantum properties of the electromagnetic field in an arbitrary, nonlinear medium. In particular, we have presented how this method can be used to describe second order nonlinear processes, and that the quantity of interest in this case is the biphoton propagator defined in Eq. \eqref{biphotonProp}. Moreover, we have presented two examples of application of our formalism, one to the very simple, and well known, case of SPDC from a one dimensional nonlinear crystal, and the other based on the origin of squeezing from multiple cascaded SPDC events.

In future works, we intend to refine this formalism, by formalising the fact, that the nonlinear interaction of the electromagnetic field in a $\chi^{(2)}$-medium under the undepleted pump approximation, can be seen as the effective free propagation of a suitable non-abelian field. Moreover, we intend to extend our results to the case of third order nonlinearities, as well as to include the quantum effects of matter, by studying photon-polariton interacitons. The model developed in this work, in fact, already contains informations about polaritons in the medium, as it has been pointed out already in Ref. \cite{Bechler}. A more detailed study of the interaction, at a quantum level, of photons and polaritons in arbitrary media, moreover, could shine new light on the origin of nonlinear effects in complex media, such as, for example, metamaterials, and metasurfaces.

\section*{Acknowledgements}
MD thanks the University of Rostock for warm hospitality. The work is part of the Academy of Finland Flagship Programme, Photonics Research and Innovation (PREIN), decision 320165.
\section*{Appendix A: Lagrangian densities of the free fields}
In this appendix, we report the explicit expressions for the free terms of the Lagrangian density appearing in Eq. \eqref{eq3}., namely the Lagrangian density of the free electromagentic field
\beq
\mathcal{L}_{em}[\vett{A}]=\frac{\varepsilon_0}{2}\dot{\vett{A}}^2-\frac{1}{2\mu_0}\left(\nabla\times\vett{A}\right)^2,
\eeq
the Lagrangian density of the free matter polarisation field
\beq
\mathcal{L}_{mat}[\vett{P}]=\frac{g(\vett{x})}{2\varepsilon_0\omega_0^2\beta(\vett{x})}\left[\dot{\vett{P}}^2-\omega_0^2\vett{P}^2\right],
\eeq
and the Lagrangian density of the reservoir field
\beq
\mathcal{L}_{res}[\vett{Y}_{\omega}]=g(\vett{x})\int_0^{\infty}\,d\omega\,\frac{\rho(\vett{x})}{2}\left[\dot{\vett{Y}}_{\omega}^2-\omega^2\vett{Y}_{\omega}^2\right].
\eeq
In the above equations, the dot indicates derivation with respect to time. Notice, moreover, that the matter polarisation field is modelled by a harmonic oscillator with resonant frequency $\omega_0$. The coefficeint $\beta(\vett{x})$ is dimensionless, and represents the static polarisability of the medium. The quantity $\rho(\vett{x})$ appearing in $\mathcal{L}_{res}$, moreover, is the mass density per unit frequency associated with each reservoir oscillator. 
\section*{Appendix B: Derivation of the effective partition function}
In this appendix, we will show, how to calculate the integrals appearing in Eq. \eqref{eq6}, using simple arguments, based on Gaussian integrals. We start by recalling the expression of a Gaussian integral in $N$ dimensions. To do that, let us consider the variable $\vett{x}=\{x_1,x_2,\cdots,x_n\}\in\mathrm{R}^n$, a constant vector $\vett{b}\in\mathbb{R}^n$, and a nonsingular $n\times n$ matrix $A\in\mathbb{C}$. Moreover, let us define the scalar product on $\mathbb{R}^n$ as $(\vett{x},\vett{y})$. The following result then holds \cite{byron}
\beq\label{eqB1}
\int\,d^nx\,e^{-\frac{1}{2}(\vett{x},A\vett{x})-(\vett{b},\vett{x})}=\sqrt{\frac{\pi^n}{\text{det}A}}e^{\frac{1}{2}(\vett{b},A^{-1}\vett{b})}.
\eeq
We can extend this results to fields, instead of vectors in $\mathbb{R}^n$, and use the above result to calculate Gaussian functional integrals involving fields. The trick to use, is to discretise the field, in such a way that, for example, the measure $\mathcal{D}\phi$ can be expressed in terms of regular integrals, i.e., $\mathcal{D}\phi\propto\Pi_{k}d\phi(\vett{x}_k)$, which can be performed using Gaussian integration in $N$ dimensions. A more complete and rigorous discussion on this method can be found, for example, in Ref. \cite{kleinert}. 

One important thing to remember, however, is that in the case of fields, the matrix $A$ appearing in the above equation, is replaced by an operator $\hat{A}$ acting on those fields. One then needs to deal with the inverse of an operator, i.e., $\hat{A}^{-1}$ to compute the integral above. It is, however, not difficult to prove, that the inverse of an operator is the Green function associated to that operator \cite{byron}. Moreover, the solution of the integral above requires, for the case of fields, the knowledge on how to calculate the determinant of an operator. For our purposes, however, this will only amount to a global normalisation constant, and we can, for the sake of simplicity, neglect it. For more informations about how to calculate the determinant of an operator, however, the reader is addressed, foe exmaple, to Ref. \cite{kleinert}.

Before proceeding with the integration, let us rewrite Eq. \eqref{eq6} in the following form, which will be easier to deal with in the next subsections:
\beq\label{eqB3}
\mathcal{Z}_{eff}[\vett{A}]=e^{\frac{i}{\hbar}S_{em}[\vett{A}]}\mathcal{I}_P[\vett{A}],
\eeq
where
\beq\label{eqB4}
\mathcal{I}_P[\vett{A}]=\int\,\mathcal{D}\vett{P}\,e^{\frac{i}{\hbar}\left(S_{mat}[\vett{P}]+S_{mf}[\vett{A},\vett{P}]\right)}\mathcal{I}_{Y}[\vett{P}],
\eeq
and
\beq\label{eqB5}
\mathcal{I}_{Y}[\vett{P}]=\int\mathcal{D}\vett{Y}_{\omega}\,e^{\frac{i}{\hbar}\left(S_{res}[\vett{Y}_{\omega}]+S_{mr}[\vett{Y}_{\omega},\vett{P}]\right)},
\eeq
and the various actions defined above are defined according to the definitions of the correspondent Lagrangian densities defined in Eq. \eqref{eq3}. 
\subsection*{B1: Calculation of $\mathcal{I}_Y[\vett{P}]$}
To calculate $\mathcal{I}_Y[\vett{P}]$, we need to rewrite it in a form similar to that of Eq. \eqref{eqB1}. To do so, we need to write the exponent in Eq. \eqref{eqB5}, i.e., 
\barr\label{eqB6}
S_{res}[\vett{Y}_{\omega}]&+&S_{mr}[\vett{Y}_{\omega},\vett{P}]=\int\,dt\,d^3x\,\int_0^{\infty}\,d\omega\,g(x)\nonumber\\
&\times&\left[\frac{\rho}{2}\dot{\vett{Y}}_{\omega}^2-\frac{\rho\omega^2}{2}\vett{Y}_{\omega}^2-f(\omega)\vett{P}\cdot\dot{\vett{Y}}_{\omega}\right],
\earr
as a quadratic form, i.e., $(\vett{Y}_{\omega},\hat{A}\vett{Y}_{\omega})$ \cite{note2}.
To bring the above term in the desired form, we can first integrate by parts, with respect to time, the last term, to shift the time derivative, from the reservoir field, to the matter field. Then, we can transform the first term by using the identity
\beq
\left(\frac{\partial\phi}{\partial t}\right)^2=\frac{\partial}{\partial t}\left(\phi\frac{\partial\phi}{\partial t}\right)-\phi\frac{\partial^2\phi}{\partial t^2},
\eeq
and then integrate once more by parts, with respect to time. We can then rearrange the result to obtain
\barr
&&S_{res}[\vett{Y}_{\omega}]+S_{mr}[\vett{Y}_{\omega},\vett{P}]=\int\,dt\,dt'\,d^3x\,\int_0^{\infty}\,d\omega\,g(x)\nonumber\\
&\times&\left[-\frac{1}{2}\left(\vett{Y}_{\omega}(t'),\hat{A}(t,t')\vett{Y}_{\omega}(t)\right)-\left(\vett{b}(t'),\vett{Y}_{\omega}(t)\right)\right],
\earr
where a second time integration has been included to express the operator $\hat{A}=\hat{A}(t',t)$, so that it can be interpreted as an actual propagator (or Green function), and
\bseq\label{eqB9}
\begin{align}
\hat{A}&\rightarrow\frac{i\rho g(x)}{2\hbar}\left(\frac{\partial^2}{\partial t^2}+\omega^2\right)\delta(t-t'),\\
\vett{b}&\rightarrow-\frac{i}{\hbar}g(x)f(\omega)\delta(t-t')\dot{\vett{P}}.
\end{align}
\eseq
We are now in the position to solve the integral in Eq. \eqref{eqB5} using Eq. \eqref{eqB1}, which gives
\beq\label{eqB10}
\mathcal{I}_Y[\vett{P}]=\mathcal{N}_Ye^{\frac{1}{2}\left(\vett{b},\hat{A}^{-1}\vett{b}\right)},
\eeq
where $\mathcal{N}_Y$ is a normalisation constant, and 
\barr
(\vett{b}&,&\hat{A}^{-1}\vett{b})=-\frac{i}{\hbar}\Big[\int\,dt\,d^3x\,\int_0^{\infty}\,d\omega\,\frac{|f(\omega)|^2g(x)}{\rho}\vett{P}(t)^2\nonumber\\
&+&\int\,dt\,dt'\,d^3x\,\frac{g(x)}{\rho}\vett{P}(t)\mathcal{G}(t-t',x)\vett{P}(t')\Big],
\earr
where we have defined
\beq
\mathcal{G}(t-t',x)=\int_0^{\infty}\,d\omega\,\omega^2|f(\omega)|^2D_F(t-t',\omega),
\eeq
as the time-domain Green function of the reservoir field. Notice, that the inverse operator $\hat{A}^{-1}$ appearing above, can be calculated from the following solution of the one dimensional wave equation 
\beq
\left(\frac{\partial^2}{\partial\tau^2}+\omega^2\right)D_F(\tau,\omega)=\delta(\tau),
\eeq
where
\beq
D_F(t-t',\omega)=\int\,\frac{d\Omega}{2\pi}\,\frac{e^{i\Omega(t-t')}}{\omega^2-\Omega^2},
\eeq
is the Feynman propagator \cite{qft}. With this in mind, it is then not difficult to show, that
\beq
\hat{A}^{-1}\rightarrow\frac{\hbar}{i\rho g(x)}D_F(t-t',\omega).
\eeq
\subsection*{B2: Calculation of $\mathcal{I}_P[\vett{A}]$}
We can now turn our attention on the integral in Eq. \eqref{eqB4}, given the results we obtained above for $\mathcal{I}_Y[\vett{P}]$. The method to solve this integral, is pretty much the same, as the one outlined above, and it consists in first rewriting $\mathcal{I}_P[\vett{A}]$ in the form of a Gaussian integral, and then using Eq. \eqref{eqB1} to calculate it. To this aim,  let us first notice, that the exponent of Eq. \eqref{eqB10} contains a term proportional to $\vett{P}^2$, which can be summed with the correspondent quadratic term appearing in the free part of the matter action $S_{mat}[\vett{P}]$. To do this, we first define the quantity $v(\omega)=f(\omega)\sqrt{\varepsilon_0\omega_0^2\beta\rho}$, and introduce the scaled resonance frequency
\beq\label{eqB14}
\tilde{\omega}_0^2=\omega_0^2+\int_0^{\infty}\,d\omega\,\frac{|v(\omega)|^2}{\rho^2},
\eeq
so that, when summing the two terms proportional to $\vett{P}^2$ we have
\barr
&&\Big[-\frac{g(x)}{2\varepsilon_0\omega_0^2\beta}\omega_0^2-\int_0^{\infty}\,d\omega\,\frac{g(x)|f(\omega)|^2}{2\rho}\Big]\vett{P}^2\nonumber\\
&=&\frac{g(x)\tilde{\omega}_0^2}{\varepsilon_0\omega_0^2\beta}\vett{P}^2.
\earr
Essentially, Eq. \eqref{eqB14} describes a shift in the material resonance frequency, due to the presence of absorption.

The next step is then to integrate by parts the term proportional to $\dot{\vett{P}}^2$, and introduce an extra time-integration, so that the total exponent appearing in the integral \eqref{eqB4} can be then written as $(\vett{P},\hat{A}\vett{P})+(\vett{b},\vett{P})$, where in this case
\barr
\hat{A}&\rightarrow&\frac{ig(x)}{\hbar\varepsilon_0\omega_0^2\beta}\left(\frac{\partial^2}{\partial t^2}+\tilde{\omega}_0^2\right)\delta(t-t')\nonumber\\
&-&\frac{ig(x)}{\hbar\rho}\mathcal{G}(t-t',x),\\
\vett{b}&=&-\frac{ig(x)}{\hbar}\dot{\vett{A}}.
\earr
We can now carry out the integration with respect to the matter degrees of freedom $\vett{P}$ using Gaussian integration. This gives the following result
\beq
\mathcal{I}_P[\vett{A}]=e^{\frac{i}{2\hbar}\int\,dt\,dt'\,d^3x\,g(x)\,\dot{\vett{A}}(x,t)\Gamma(t-t',x)\dot{\vett{A}}(x,t')},
\eeq
where $\Gamma(t-t',x)$ is the solution of the following, integro-differential equation \cite{Bechler}
\barr\label{integroDifferential}
&&\frac{1}{\varepsilon_0\omega_0^2\beta}\left(\frac{\partial^2}{\partial t^2}+\tilde{\omega}_0^2\right)\Gamma(t-t',x)\nonumber\\
&-&\frac{1}{\rho}\int\,d\tau\,\mathcal{G}(t-\tau,x)\Gamma(\tau-t,x)=\delta(t-t').
\earr
\subsection*{B3: Physical meaning of $\Gamma(t-t',x)$}
Equation \eqref{integroDifferential} is, in general, very complicated, and does not admit analytical solutions. However, the function $\Gamma(t-t',x)$ contains all the informations about the medium, in which the electromagnetic field is propagating, and can be interpreted, as a kind of effective dielectric constant for the ``dressed" electromagnetic field. This, moreover, is corroborated by the calculations presented in Ref. \cite{Bechler}, where it is shown, that, essentially, $\Gamma(t-t',x)$ can be interpreted as an effective dielectric constant. To understand this, let us assume, that $\tilde{\Gamma}(\Omega,x)$ is the Fourier transform of $\Gamma(t-t',x)$. The effective Lagrangian associated  to the effective action defined in Eq. \eqref{effectiveAction}, can be written, in terms of the electric and magnetic fields in Fourier domain as follows
\barr
\mathcal{L}_{eff}[\vett{A}]&=&\varepsilon_0|\vett{E}(\Omega,x)|^2+\frac{1}{\mu_0}\vett{B}(\Omega,x)|^2\nonumber\\
&+&g(x)\Omega^2\tilde{\Gamma}(\Omega,x)|\vett{E}(\Omega,x)|^2.
\earr
According to standard field theory \cite{classicalMechanics}, the electric displacement $\vett{D}(\Omega,x)$ can be directly derived from the effective Lagrangian as follows:
\barr
\vett{D}(\Omega,x)&=&\frac{\partial\mathcal{L}_{eff}}{\partial\vett{E}^*}=\varepsilon_0\vett{E}(\Omega,x)\nonumber\\
&+&g(x)\tilde{\Gamma}(\Omega,x)\vett{E}(\Omega,x).
\earr
From the above equation, and by recalling the constitutive relation $\vett{D}=\varepsilon_0\varepsilon\vett{E}$ \cite{jackson}, it is possible to define the (positive frequency) effective dielectric constant as
\beq
\varepsilon_+(\Omega,x)=1+\frac{g(x)}{\varepsilon_0}\tilde{\Gamma}(\Omega,x),
\eeq
while the negative frequency part of the dielectric constant can be obtained by analytic containuation, namely $\varepsilon_-(\Omega,x)=\varepsilon_+^*(\Omega,x)$. The result above, whose derivation is given in great detail in Ref. \cite{Bechler}, is consistent with standard results in electromagnetic theory \cite{jackson}, and therefore means, that essentially the effect of the reservoir, and the matter polarisation of the medium, in which the electromagnetic field propagates, can be globally described by the dielectric function of the material itself.
\subsection*{Appendix C: Explicit expression of the $\hat{R}$ operator}
The differential operator $\hat{R}$ introduced in Sect. \ref{section3} to express the action $S_q[\vett{A},\vett{J}]$ in a quadratic form, is a vector operator, which can be written as $\hat{R}=\hat{R}^{(0)}-+g(x)\partial^2\hat{\Gamma}/\partial t^2$, where $\hat{\Gamma}$ is the operator, whose Green function is given by $\Gamma(t-t',x)$, as defined in Eq. \eqref{integroDifferential}, and $\hat{R}^{(0)}$ is a vectorial operator, whose components $\hat{R}^{(0)}_{\mu\nu}$  are given as follows:
\barr
&&\hat{R}^{(0)}_{\mu\nu}(t-t',x-x')=\Big[\Big(\varepsilon_0\frac{\partial^2}{\partial t^2}-\frac{1}{\mu_0}\nabla^2\Big)\delta_{\mu\nu}\nonumber\\
&+&\frac{1}{\mu_0}\frac{\partial}{\partial x_{\mu}}\frac{\partial}{\partial x_{\nu}}\Big]\delta(t-t')\delta(x-x'),
\earr
where $(\mu,\nu)\in\{1,2,3\}$, and we set $\{x,y,z\}\equiv\{x_1,x_2,x_3\}$.


\begin{thebibliography}{99}
\bibitem{note} If the medium is homogeneous, the spectral coupling $f(\omega)$ (as well as the polarisability $\beta$, and the mass density per unit frequency $\rho$, appearing in the expressions of the free Lagrangian densities) is constant with respect to $\vett{x}$.
\bibitem{note3} A more formal approach to this calculation will involve the proper adaptation of the Faddeev-Popov technique, to the case of non-relativistic fields, and use it to demonstrate, that our result is correct, within the framework of the Weyl gauge. We leave the discussion of this method to a future work.
\bibitem{note4} Notice, moreover, that the constant term, corresponding to $P=0$ has been omitted, since it is of no interest in this analysis.
\bibitem{note2} For the sake of simplicity, we implicitly defined $x\equiv\vett{x}$, $\rho\equiv\rho(\vett{x})$, and $f(\omega)\equiv f(\omega,\vett{x})$.
\end{thebibliography}
\end{document}